\documentclass[11pt,a4paper]{article}

\usepackage{amsfonts}
\usepackage{graphicx}
\usepackage[colorlinks=true,linkcolor=blue]{hyperref}

\title{\textbf{One-loop kink mass shifts: \\ A computational approach.}}

\author{A. Alonso Izquierdo$^{(a)}$, and J. Mateos Guilarte$^{(b)}$
\\ {\normalsize {\it $^{(a)}$ Departamento de Matematica
Aplicada and IUFFyM}, {\it Universidad de Salamanca, SPAIN}} \\ {\normalsize {\it $^{(a)}$ Departamento de Fisica
Fundamental and IUFFyM}, {\it Universidad de Salamanca, SPAIN}}}

\date{}

\begin{document}

\maketitle

\begin{abstract}
In this paper we develop a procedure to compute the one-loop quantum correction to the kink masses in generic (1+1)-dimensional one-component scalar field theoretical models. The procedure uses the generalized zeta function regularization method helped by the Gilkey-de Witt asymptotic expansion of the heat function via Mellin's transform. We find a formula for the one-loop kink mass shift that depends only on the part of the energy density with no field derivatives, evaluated by means of a symbolic software algorithm that automates the computation. The
improved algorithm with respect to earlier work in this subject has been tested in the sine-Gordon and $\lambda(\phi)_2^4$ models. The quantum corrections of the sG-soliton and $\lambda(\phi^4)_2$-kink masses have been estimated with a relative error of $0.00006\%$ and $0.00007\%$ respectively. Thereafter, the algorithm is applied to
other models. In particular, an interesting one-parametric family of double sine-Gordon models interpolating between the ordinary sine-Gordon and a re-scaled sine-Gordon model is addressed. Another one-parametric family, in this case of $\phi^6$ models, is analyzed. The main virtue of our procedure is its versatility: it can be applied to practically any type of relativistic scalar field models supporting kinks.
\end{abstract}


\section{Introduction}

In this paper we shall address the broad topic of one-loop quantum corrections to kink masses in (1+1)-dimensional real scalar field theory models. Kinks are solitary waves that arise in non-linear models. This kind of solution resembles at the classical level an extended particle, i.e., they are localized, finite-energy objects. Some particles in nature, mostly baryons, seem to be extended, although they are sometimes approximated by point particles in theoretical calculations. In 1961 Skyrme pointed out that, in a certain extension of the non-linear sigma model now called the Skyrme model, there exist both 3D dispersive and solitary waves among their solutions \cite{Skyrme}. Because the model attempted to describe low-energy hadron phenomenology, and because solitary waves are formed
from a heavy classical lump of energy, the idea is natural: upon quantization, dispersive waves become light mesons -pions- and solitary waves give rise to heavy baryons -protons, neutrons-.
This bold idea prompted the task of investigating solitary waves in the quantum domain, a task mainly performed in the seventies, see e.g. \cite{Dashen1974,Faddeev1978,Cahill1976}.

The first theoretical studies in this issue were accomplished by Dashen, Hasslacher and Neveu by studying the
quantum $\lambda (\phi)^4_2$ and sine-Gordon kinks. In 1974 they
succeeded in computing the one-loop
correction to the classical mass of these solitary waves by
developing the $\hbar$-expansion of these (1+1)-dimensional field
theories. From the work of these authors in papers \cite{Dashen1974,Dashen1975}  emerged the Dashen-Hasslacher-Neveu (DHN) formula{\footnote{To be precise, the formula for static kinks should be called the first DHN formula. There is a second DHN formula that calibrates the WKB correction to the Bohr-Sommerfeld levels of quantum breathers, see \cite{Coleman1985}. }} for the kink mass quantum correction.  The DHN formula demands knowledge of the spectrum of the variational second-order (Hessian) differential operators valued at the kink/soliton solutions. In the s-G and $\lambda (\phi)^4_2$ models these operators are Schr$\ddot{\rm o}$dinger operators of P$\ddot{\rm o}$schl-Teller type, whose spectral problem is solvable. Moreover, the second-order fluctuation operators around the sG and $\lambda\phi^4$ kinks are respectively the first and the second of P$\ddot{\rm o}$sch-Teller Hamiltonians in the hierarchy of transparent -
reflection coefficient equal to zero- models of this type. This provides sufficient spectral information to
achieve exact computation of the one-loop quantum correction to the kink masses through the DHN formula: 1) $\Delta E_{sG}(\phi_K)=-\hbar \frac{m}{\pi}$, \, 2) $\Delta E_{\lambda\phi^4}(\phi_K)= \hbar \frac{m}{\sqrt{2}}\left( \frac{1}{2\sqrt{3}}-\frac{3}{\pi} \right)$.

Apart from these two models the DHN formula has little computational value. The reason is that the Hessian differential operator valued on the kink solution that emerges in other models has an unknown spectrum in general. Therefore, computation of the one-loop quantum correction to the kink mass in a generic model required the
development of a new procedure; see the Reference \cite{Mateos2009} to find a summary of the work of our group on this issue during the last ten years. In fact, we started the new approach simultaneously with renewed interest about the problem of the quantization of classical lumps at the end of the last century. The new impetus came from the quantization of supersymmetric kinks, where the contributions associated with fluctuations of the bosonic and the fermionic fields balance each other, which avoids the need to identify the spectrum of the Hessian operators involved in this problem. However, new subtleties concerning the effect of the boundary conditions come into play. Several groups at Stony Brook/Wien, \cite{Rebhan1997,Nastase1999}, Minnesota \cite{Shifman1999}
and MIT \cite{Graham1999,Graham1999b} addressed mixed issues in the problem by studying the impact made by using different types of boundary conditions - PBC, Dirichlet, Robin-, regularization
methods -energy cutoff, mode number cutoff, high-derivatives-, and/or performed phase shift analysis,
in connection with possible modifications due to the quantum effects of the central charge
of the SUSY algebra \cite{Bordag1995,Bordag2002}. The Stony Brook/Wien group formed by
Rebhan, van Nieuwenhuizen, and Wimmer, together with Goldhaber, investigated the
computations of mass shifts induced by one-loop fluctuations on
supersymmetric kinks \cite{Wimmer2001,Goldhaber2004,Rebhan2004}. The extreme elusiveness of this issue did not
prevent those authors from identifying the old DHN formula as being
based on a regularization method that sets a cutoff in the number of
fluctuation modes to be counted, rather than the conventional energy
cutoff. Another group from Minnesota University addressed the same
problem by using high-derivative regularization, with SUSY being
preserved by boundary conditions to find similar results.
Phase-shift analysis by an MIT group -Jaffe, Graham, and
collaborators- also led to some advances, in this case in a purely
bosonic setting  \cite{Graham1999}.

Our goal in this paper is threefold. First, we construct a better structural derivation of the DHN formula and provide the one-loop kink mass shifts of the sG- and $\lambda\phi^4$-kinks after a regularization by means of a cutoff in the number of modes. The same results are achieved by a treatment of the DHN formula based on the heat kernel/zeta function regularization of ultraviolet divergences.

Second, the method of zeta function regularization was invented by Dowker and Critchley and, independently by Hawking, circa 1976 \cite{Hawking1977}. The partition function of Euclidean quantum field theories is a functional integral that, up to one-loop order in the
$\hbar$-expansion, is the inverse of the square root of the
determinant of a differential operator of Laplace type times
the exponential of the classical Euclidean action over $\hbar$.
In our framework, the pertinent differential operators
are those ruling small quantum fluctuations in the classical kink backgrounds. Generically, the
spectral information in these situations is grossly insufficient for
identifying the generalized zeta function in terms of known spectral
functions. Fortunately, in
the mid sixties de Witt \cite{DeWitt1965} had already proposed the use of the high-temperature expansion of the kernel
of the generalized heat equation related to the differential
operator of Laplace type to unveil the meromorphic structure of the
generalized zeta function. To achieve this goal, one takes advantage
of the link between generalized heat and zeta functions via Mellin
transforms, such that the residua at the poles of the generalized
zeta function are proportional to the Seeley coefficients of the
heat kernel expansion \cite{Gilkey1984, Roe1988, Elizalde1994, Kirsten2002, Vassilevich2003, Alonso2002, Alonso2002A}. We shall use all this theoretical machinery in order to construct a formula for the one-loop quantum correction to the kink mass in a generic (1+1)-dimensional real scalar field theory as a truncated series in the Seeley coefficients
of the Hessian heat function.

Third, the effectiveness of this method relies on the computation of the Seeley coefficients. This is not an easy task and requires the solution of difficult recurrence relations between the Seeley densities and integration of these densities over the whole real line. In this paper we shall improve our symbolic method implemented on a PC computer
in a Mathematica environment in two ways: 1) We shall adapt in an optimum way the range of the inverse temperature in the integral Mellin transform  to the number of Seeley coefficients kept in the truncated formula. 2) We shall use the first-order equations to skip the integration over the whole spatial line traded by integration in the field space
between the two vacua connected by the kink. These two advances offers not only great computational advantage but also
higher precision. The error in the computation of the one-loop sG-kink is diminished from $6.0\%$ to the $0.00006\%$ and for the $\lambda\phi^4$-kink the reduction is from $0.07\%$ to the $0.00007\%$ !!! But this is not all: several other models for which the spectral information is not available can now be treated and the one-loop
mass shift calculated. Among them the double sine-Gordon model \cite{Mussardo, Bazeia}, the Razavy potential
\cite{Razavy}, the BGLM model \cite{BGLM} and others. Especially interesting is the application of the method to a family of trigonometric models interpolating between the sine-Gordon and the re-scaled sine-Gordon models. A family of deformations of the $\phi^6$ model \cite{Lohe, Khare} will be also analyzed.

The organization of this paper is as follows. In Section \S. 2 and \S. 3 we shall review the standard lore about kink topological defects and their one-loop fluctuations. This review will allow us to fix the notation and to remark on some important features of this kind of solutions. Section \S. 4 will be devoted to the DHN formula.  A cutoff in the number of modes and, alternatively, the generalized or spectral zeta function procedure will be used in the regularization of intermediate divergent quantities. We shall apply this formula to the $\phi^4$ and sine-Gordon model examples. In section \S. 5 the Seeley-Gilkey-deWitt-Abramidi heat kernel expansion will be introduced to express the generalized zeta function as an asymptotic series. We stress that the computation of the Seeley coefficients only requires the form of the potential. Again, we address the $\lambda\phi^4$ and sine-Gordon model examples in order to check the accuracy of these formulas. Section \S. 6 is devoted to obtaining the asymptotic formula to be summarized in a symbolic algorithm, whereas in Section \S. 7 we construct an efficient program for that algorithm. We test the algorithm on the $\phi^4$ and sine-Gordon models, gaining a feedback about the method.  Finally, in Section \S.8
we shall apply the program to several interesting scalar field theory models and families of models.

\section{Classical field theoretical models}

The action governing the dynamics in our (1+1)-dimensional relativistic one-scalar field theoretical models is of the form:
\[
\tilde{S}[\psi]=\int \!\! \int\, dy^0dy^1 \, \left(\frac{1}{2}\frac{\partial\psi}{\partial y_\mu}\cdot \frac{\partial\psi}{\partial y^\mu}- \tilde{U}[\psi(y^\mu)] \right) \quad .
\]
Here, $\psi(y^\mu): \mathbb{R}^{1,1} \rightarrow \mathbb{R}$ is a real scalar field; i.e., a continuous map from the $1+1$-dimensional Minkowski space-time to the field of the real numbers.  $y^0=\tau $ and $y^1=y$ are local coordinates in ${\mathbb R}^{1,1}$, which is equipped with a metric tensor $g_{\mu\nu}={\rm diag}(1,-1), \mu,\nu=0,1$ such that $y_\mu y^\mu=g^{\mu\nu}y_\mu y_\nu$, $\frac{\partial}{\partial y_\mu}.\frac{\partial}{\partial y^\mu}=g^{\mu\nu}\frac{\partial}{\partial y_\mu}.\frac{\partial}{\partial y_\nu}$.

We shall work in a system of units where the speed of light is set to one, $c=1$, but we shall keep the Planck constant $\hbar$ explicit because we shall search for one-loop corrections, proportional to $\hbar$, to the classical kink masses.
In this system, the physical dimensions are:
\[
[\hbar]=[\tilde{S}]=M L \quad , \quad [y_\mu]=L \quad , \quad [\psi]=M^\frac{1}{2}L^\frac{1}{2} \quad , \quad [\tilde{U}]=ML^{-1} \quad .
\]
The models that we shall consider are distinguished by different choices of the part of the potential energy density which is independent of the field spatial derivatives: $\tilde{U}[\psi(y^\mu)]$. In all of them there will be two special parameters, $m_d$ and $\gamma_d$, to be determined in each case, carrying the physical dimensions dimensions: $[m_d]=L^{-1}$ and $[\gamma_d]=M^{-\frac{1}{2}}L^{-\frac{1}{2}}$. We define the non-dimensional coordinates, fields and potential in terms of these parameters:
\[
x_\mu=m_d y_\mu \, \, \, , \, \, \, x_0=t \, \, , \, \, x_1=x \quad , \quad \phi= \gamma_d \psi \qquad , \qquad U(\phi)=\frac{\gamma_d^2}{m_d^2} \tilde{U}(\psi) \qquad .
\]
The action is also proportional to a non-dimensional action
\begin{equation}
\tilde{S}[\psi]=\frac{1}{\gamma_d^2}S[\phi]= \frac{1}{\gamma_d^2}\int\!\!\int\, dx^0dx^1 \, \left(\frac{1}{2}\frac{\partial\phi}{\partial x_\mu}\cdot \frac{\partial\phi}{\partial x^\mu}- U[\phi(x^\mu)] \right) \label{action}
\end{equation}
and the field equation in this non-dimensional setup is the PDE:
 \begin{equation}
\frac{\partial^2 \phi}{\partial t^2} - \frac{\partial^2 \phi}{\partial x^2} = -\frac{\partial U}{\partial \phi}\quad .
\label{pde}
\end{equation}
 Unless $U(\phi)$ is a quadratic polynomial in the field, the PDE (\ref{pde}) is non-linear. Moreover, we shall assume that $U(\phi)$ is a non-negative twice-differentiable function of $\phi$: $U(\phi)\in C^2({\mathbb{R}})$ and $U(\phi)\geq 0$ for $\phi \in \mathbb{R}$.
\subsection{The configuration space and the vacuum orbit}
The configuration space ${\cal C}$ of the system is the set of field configurations -the continuous maps from $\mathbb{R}$ to $\mathbb{R}$ for fixed time $t=t_0$- for which the potential energy functional
\begin{equation}
\tilde{E}[\psi]=\frac{m_d}{\gamma_d^2}E[\phi]=\frac{m_d}{\gamma_d^2}\int_{-\infty}^\infty \, dx \, \left[ \frac{1}{2} \left( \frac{\partial \phi}{\partial x} \right)^2 + U(\phi) \right]
\label{energy}
\end{equation}
is finite. Thus, ${\cal C}=\{\phi(t_0,x)\in {\rm Maps}(\mathbb{R}^{1},\mathbb{R})/ E[\phi]<+\infty\}$. Let
\begin{equation}
{\cal M}=\{\phi^{(i)}\,\,/ \,\,U(\phi^{(i)})=0\}
\label{hsol}
\end{equation}
be the set of zeroes of $U$, which we assume to be a finite or infinite discrete set. The initial conditions $\phi(0,x)=\phi^{(i)}$, $\dot{\phi}(0,x)=0$ provide the static and homogeneous solutions to the PDE (\ref{pde}).
When $i>1$, a process of spontaneous symmetry breaking is triggered by the choice of vacuum in the
quantization of these classical systems. In our models there will be a discrete symmetry group $G$ and the set ${\cal M}$ is the union of one or several orbits of the action of $G$ on any of the vacua: ${\cal M}=\bigsqcup_A \frac{G}{G_A}$, where $G_A, A=0,1, \cdots , r$, is the little group in the $A$ orbit. The vacuum moduli space is in turn the set of orbits: ${\cal N}=\frac{{\cal M}}{G}=\{\phi^{(0)}, \phi^{(1)}, \cdots , \phi^{(r)}\}$.

Each field configuration in ${\cal C}$ must comply with the asymptotic conditions
\begin{equation}
\lim_{x\rightarrow \pm \infty} \phi(t,x)\in {\cal M} \hspace{0.5cm},\hspace{0.5cm} \lim_{x\rightarrow \pm \infty} \frac{\partial \phi(t,x)}{\partial x}=0.
\label{asymptotic}
\end{equation}
in order to guarantee the finiteness of the energy (\ref{energy}). Thus, the configuration space is the union of topologically disconnected sectors
\[
{\cal C}=\cup_{i,j}{\cal C}_{ij}
\]
where ${\cal C}_{ij}$ stands for the set of the configurations that asymptotically connects the element $\phi^{(i)}$ with the element $\phi^{(j)}$ in ${\cal M}$, i.e., we assume that:
\[
\lim_{x\rightarrow -\infty} \phi(t,x)=\phi^{(i)}\in {\cal M} \hspace{0.5cm}, \hspace{0.5cm} \lim_{x\rightarrow +\infty} \phi(t,x)=\phi^{(j)}\in {\cal M}
\]
Because temporal evolution is continuous (a homotopy transformation), the asymptotic conditions do not change in time and the sectors ${\cal C}_{ij}$ are topologically disconnected. Note that the elements of ${\cal M}$ belong to the ${\cal C}_{ii}$ sectors.

\subsection{Kinks and solitary waves}

In the ${\cal C}_{i\, i+1}$ sectors there cannot exist static homogeneous solutions. In these sectors there are, however, spatially dependent static solutions of (\ref{pde}) of solitary wave type: non dispersive non-linear  waves, see e.g. \cite{Rajaraman1982, Drazin1996, Manton2004}. For static field configurations the PDE (\ref{pde}) reduces to the ODE:
\begin{equation}
\frac{d^2 \phi}{d x^2} = \frac{\partial U}{\partial \phi} \label{ode}
\end{equation}
 The ODE (\ref{ode}) can be seen as the equation of motion of a mechanical system: understand $\phi$ as the particle coordinate, $x$ as the particle time, and $-U$ as the particle potential energy. The second order ODE (\ref{ode}) admits the first-integral
\[
I=\frac{1}{2} \left( \frac{d\phi}{dx} \right)^2 - U(\phi) \quad ,
\]
which is the particle energy in the analogous mechanical system. The asymptotic conditions (\ref{asymptotic}) guaranteeing finite field theoretical energy -particle action- force zero energy $I=0$ and the solitary wave solutions satisfy the first-order ODE
\begin{equation}
\frac{d\phi}{dx} = \pm\sqrt{2 U(\phi)} \qquad .
\label{ode1}
\end{equation}
Thus, the solitary waves are in one-to-one correspondence with the separatrix trajectories between the bound and unbound motions of the mechanical system connecting contiguous unstable equilibrium points. In fact, the zero mechanical energy Hamilton characteristic function
\[
W(\phi)=\pm \int \, d\phi \, \sqrt{2U(\phi)}
\]
provides the particle action -henceforth the solitary wave or kink energy - as a topological quantity:
\[
E(\phi_K)=\left|W(\phi^{(i)})-W(\phi^{(i+1)})\right| \quad .
\]
Moreover, a Lorentz transformation sends the static solution $\phi_K(x)$ -in the kink center of mass- to another time-dependent solution $\phi_K(t,x)=\phi_K(\frac{x-v t}{\sqrt{1-v^2}})$, showing the kink as a traveling wave.

\subsection{The prototypes: sine-Gordon and $\phi^4$ kinks}

We now collect very well known facts about these structures in the prototypical
sine-Gordon and $\phi^4$ models for the sake of completeness.

\subsubsection{The sine-Gordon model}

In this famous model, the dimensionless potential in (\ref{action}) is the trigonometric function
\[
U(\phi)=1-\cos \phi
\]
whereas the special parameters are $m_d^2=m^2$ and $\gamma_d^2=\frac{\lambda}{m^2}$. Therefore, ${\cal M}=\{\phi^{(n)}=2\pi n, n\in{\mathbb Z}\}$. The symmetry group is $G= {\mathbb Z}_2\times {\mathbb Z}={\mathbb D}_\infty$, the infinite dihedral group encompassing the ${\mathbb Z}_2$ group generated by reflection in the fields $\phi\to -\phi$ times field translations in $2\pi n$: $\phi\to\phi+2\pi n \, , \, n\in{\mathbb Z}$. Thus, the set of zeroes of $U$ is the orbit of the symmetry group, ${\cal M} = \frac{{\rm G}}{{\rm G}_0}$, where in this case $G_0={\mathbb Z}_2$. In our context, the vacuum moduli space is the quotient space ${\cal N}=\frac{{\cal M}}{G}$ and it is a single point because the action of $G$ on ${\cal M}$ is transitive, meaning that all the elements in ${\cal M}$ are equivalent ${\cal N}=\{\phi^{(0)}\}$. Finally, the soliton or kink solution in the topological sector ${\cal C}_{n\,n\pm 1}$ of (\ref{ode1}) is:
\[
\phi_K(x)=\pm 4 \arctan e^{x-x_0}+2\pi n \quad .
\]
The zero mechanical energy Hamilton characteristic function $W(\phi)=\pm 4 \cos \frac{\phi}{2}$
 leads us to the kink classical energy: $E(\phi_K)=4\left|{\rm cos}\frac{\phi^{(n)}}{2}-{\rm cos}\frac{\phi^{(n+1)}}{2} \right|=8$.

\subsubsection{The $\lambda\phi^4$ model.}

In this model the dimensionless potential in (\ref{action}) is a polynomial:
\[
U(\phi)=\frac{1}{2}(\phi^2-1)^2
\]
and the special parameters are $m_d^2=\frac{m^2}{2}$ and
$\gamma_d^2=\frac{\lambda}{m^2}$. The set of zeroes of the potential is ${\cal M}=\{\phi^{(i)}=(-1)^i, i=0,1\}$. The symmetry group is $G\simeq\{\phi\to\phi, \phi\to -\phi\}={\mathbb Z}_2$, the discrete group generated by reflection in the fields. The set of zeroes of $U$ is the orbit of the symmetry group $\frac{{\rm G}}{{\rm G}_0}={\cal M}$ where $G_0={\mathbb I}$ is the identity. The \lq\lq moduli space of vacua" is again a one-element set, and the quotient space is: ${\cal N}=\frac{{\cal M}}{G}= \{\phi^{(0)}\}$. The kink solutions of (\ref{ode1}) are now:
\[
\phi_K(x)=\pm {\rm tanh}(x-x_0) \quad .
\]
The zero mechanical energy Hamilton characteristic function is $W(\phi)=\pm \left(\frac{\phi^3}{3}-\phi\right)$. Therefore, the kink classical energy is: $E(\phi_K)=\left|\frac{2}{3}-2\right|=\frac{4}{3}$.

\section{One-loop quantum fluctuations}

\subsection{Vacuum fluctuations}

The small (quadratic) fluctuations around any of the equivalent constant solutions $\phi(t,x)=\phi^{(i)}+\delta\phi(t,x)$ satisfy the linearized field equations:
\begin{equation}
\left(\frac{\partial^2}{\partial t^2}-\frac{\partial^2}{\partial x^2}+v^2\right)\delta\phi(t,x)+{\cal O}[(\delta\phi)^2]=0 \quad , \quad \left.\frac{\partial^2 U}{\partial\phi^2}\right|_{\phi^{(i)}}=v^2 \quad . \label{lpde}
\end{equation}
The solutions of (\ref{lpde}) of the form $\delta\phi_k(t,x)=e^{i\nu(k)t}f_k(x)$ are the vacuum normal modes of fluctuation coming from the eigenfunctions of the $K_0=-\frac{d^2}{dx^2}+v^2$ differential operator:
\[
K_0f_k(x)=\nu^2(k)f_k(x)\quad , \quad \nu^2(k)=k^2+v^2 \quad , \quad f_k(x)=e^{ikx} \quad , \quad k\in{\mathbb R} \quad .
\]
In order to tame the difficulties of dealing with a continuous spectrum, we place the system in a normalization interval of very large, but finite, non-dimensional length, $l=m_d L$, and choose periodic boundary conditions on the plane waves: $f_k(-\frac{l}{2})=f_k(\frac{l}{2})$. The spectrum of $K_0$ becomes discrete and the eigenfunctions are monochromatic plane waves:
\[
f_{k_n}(x)=\frac{1}{\sqrt{l}}e^{ik_n x}\quad , \quad k_n=\frac{2\pi}{l}n \, \, , \, \, n\in{\mathbb Z} \quad , \quad \nu^2(k_n)=\frac{4\pi^2}{l^2}n^2+v^2 \quad .
\]
Therefore, the general solution of (\ref{lpde}) with PBC is a Fourier series, the more general linear combination of these normal modes. In quantum theory, Fourier coefficients become creation and annihilation operators of the fundamental mesons moving in the vacuum, and the energy of the vacuum state, when all these fluctuation states are unoccupied, is the one-loop shift to the vacuum energy:
\[
\bigtriangleup\tilde{E}_0[\psi^{(i)}]=\frac{m_d}{\gamma_d^2}\bigtriangleup E_0[\phi^{(i)}]=\frac{\hbar m_d}{2}\left(\lim_{N\to\infty}\sum_{-N}^N\, \nu(k_n)\right) \quad .
\]
Putting the system in an interval with periodic boundary conditions, the vacuum spectrum becomes discrete and setting an integer number $N$ to be large, but finite, a cutoff in the energy of the fluctuation modes is selected. At the end of the day, however, we must come back to the $l\to\infty$ and $N\to\infty$ limits. Bearing in mind the spectral density of the fluctuation modes $\rho_0(k)=
\frac{dn}{dk}=\frac{l}{2\pi}$, we have the following expression for the vacuum energy at this double limit:
\[
\bigtriangleup E_0[\phi^{(i)}]=\frac{\hbar\gamma_d^2}{2}\left(\int_{-\infty}^\infty \, dk \, \rho_0(k)\nu(k)+\frac{v}{2}\right)=\frac{\hbar\gamma_d^2}{2}\left(\frac{l}{2\pi}\int_{-\infty}^\infty \, dk \, \sqrt{k^2+v^2} +\frac{v}{2}\right) \quad .
\]
We now remark an obscure and subtle point. At $l=\infty$, a half-bound state with energy $\nu_b=v$ and eigenfunction constant emerges. This state is different from the $\nu_{k=0}=v$ state, the threshold of the continuous spectrum. The name is suggested by the fact that a factor multiplies the eigenvalue of the half-bound state in its contribution to the vacuum energy by one-half because of the one-dimensional Levinson theorem. To see a proof of this delicate point, see the APPENDIX in \cite{Alonso2004} which is worked following the seminal paper of Barton \cite{Barton}. We also notice that in the $l\to\infty$ limit one goes beyond the Hilbert space $L^2({\mathbb R})$ to its completion $\bar{L}^2({\mathbb R})$ by admitting monochromatic plane waves that are not square-summable.

\subsection{Kink fluctuations}

The small fluctuations over any of the equivalent kink solutions $\phi(t,x)=\phi_K(x)+\delta\phi(t,x)$ satisfy the linearized field equations:
\begin{equation}
\left(\frac{\partial^2}{\partial t^2}-\frac{\partial^2}{\partial x^2}+v^2+V(x)\right)\delta\phi(t,x)+{\cal O}[(\delta\phi)^2]=0 \quad , \quad \left.\frac{\partial^2 U}{\partial\phi^2}\right|_{\phi_K(x)}=v^2+V(x) \quad . \label{lpdek}
\end{equation}
The solutions of (\ref{lpdek}), of the form $\delta\phi_k(t,x)=e^{i\omega(q)t}f_q(x)$, are the normal modes of kink fluctuations coming from the eigenfunctions of the $K=-\frac{d^2}{dx^2}+v^2+V(x)$ differential operator:
\[
\lim_{x\rightarrow \pm \infty} V(x) =0 \quad , \quad Kf_q(x)=\omega^2(q)f_q(x) \quad , \quad  \omega^2(q)=q^2+v^2 \quad .
\]
Because of conditions (\ref{asymptotic}), the kink solutions asymptotically tend to the vacuum solutions when $x\to\pm\infty$. For this reason $V(x)$ tends to zero both at $x\to +\infty$ and at $x\to -\infty$. This is true in every model for which $\left.\frac{\partial^2 U}{\partial\phi^2}\right|_{\phi^{(i)}}=v^2=\left.\frac{\partial^2 U}{\partial\phi^2}\right|_{\phi^{(i+1)}}$, a property that always happens if the vacuum moduli space is a set of one element. In this case{\footnote{In fact, $\lim_{x\to +\infty}\, K=K_{i+1}=-\frac{d^2}{dx^2}+v_{i+1}^2$, $\lim_{x\to -\infty}\, K=K_{i}=-\frac{d^2}{dx^2}+v_{i}^2$ if $\left.\frac{\partial^2 U}{\partial\phi^2}\right|_{\phi^{(i)}}=v_i^2 \neq \left.\frac{\partial^2 U}{\partial\phi^2}\right|_{\phi^{(i+1)}}=v_{i+1}^2$. This happens when the kink interpolates between different points of the vacuum moduli space: $\phi^{(i)}$ and $\phi^{i+1}$.}},
\[
\lim_{x\to\pm\infty}K=K_0 \qquad .
\]
Unlike $K_0$, however, which is a Helmoltz operator, $K$ is an Schr$\ddot{\rm o}$dinger operator. The kink normal modes are not plane waves but some dispersive wave functions distorted by the kink. Bound states and half-bound states can also exist. Thus, there are three types of eigenfunctions of $K$ at the $l=\infty$ limit that are characterized as follows:
\begin{enumerate}

\item Scattering states. The asymptotic behaviour is:
\[
f_q(x)\stackrel{x\to\pm\infty}{\simeq} {\rm exp}[iqx+\frac{1}{2}\delta(q)] \qquad , \qquad q\in{\mathbb R} \quad ,
\]
where $\delta(q)$ is the total phase shift induced by $V(x)$. Periodic boundary conditions require that
\[
q_nl+\delta(q_n)=2\pi n \qquad .
\]
Therefore, $q_n=k_n-\frac{1}{l}\delta(q_n)$ and the spectral density over the kink background is: $\rho(k)=\frac{l}{2\pi}+\frac{1}{2\pi}\frac{\partial\delta(k)}{\partial k}$.

\item Bound states: $f_{\omega_1}(x)$, $f_{\omega_2}(x)$, $\cdots$ , $f_{\omega_{b}}(x)$ with discrete positive eigenvalues
$\omega_1^2=0<\omega_2^2<\omega_3^2< \cdots<\omega_{b}^2<v^2$ below the threshold of the continuous spectrum $v^2$ and classified by the number of nodes $b$. The asymptotic behaviour is:
\[
\lim_{x\to -\infty}f_{\omega_i}(x)\propto e^{(v-\omega_i)x} \quad , \quad \lim_{x\to +\infty}f_{\omega_i}(x)\propto e^{-(v-\omega_i)x}
\quad , \quad i=1,2, \cdots , b \qquad .
\]
These wave functions are strictly $L^2({\mathbb R})$.

\item A half-bound state. If the value of the last eigenvalue is $\omega_b^2=v^2$ then this eigenvalue involves the existence of a half-bound state, which is buried in the continuous spectrum threshold. This happens when the reflection scattering amplitude is zero.

\end{enumerate}
In particular, the lowest bound state in the kink sector is always a bound state. From ODE (\ref{ode}) one immediately sees that:
\[
\frac{d^2}{dx^2}\cdot\frac{d\phi_K}{dx}=\left.\frac{\partial^2 U}{\partial\phi^2}\right|_{\phi_K}\cdot\frac{d\phi_K}{dx} \quad .
\]
There is always a zero energy eigenfunction of the $K$ operator in the kink sector: $f_1(x)=\frac{d\phi_K}{dx}\, , \, \omega_1=0$. This eigenfunction is no more than the Goldstone boson due to the spontaneous symmetry breaking by the kink of the invariance of the system with respect to spatial translations: $x\to x+a$. Moreover, because $\phi_K(x)$ is a monotonic increasing (or decreasing) function of $x$, the translational zero mode $f_1(x)=\frac{d\phi_K}{dx}$ has no nodes. According to Sturm-Liouville theory $f_1(x)$ is thus the lowest energy eigenfunction - the ground state - of the non-negative Schr$\ddot{\rm o}$dinger operator $K$.

We write the general solution of (\ref{lpdek}) in terms of these kink normal modes, assuming a normalization interval of length $l$ and PBC{\footnote{Quantum field theories that are based on one-particle states containing some bound state are non-unitary. This is, however, the situation that we encounter in the kink sectors, see e.g. \cite{JMMCJM} }}:
\begin{eqnarray*}
\delta\phi(t,x)&=&\sqrt{\frac{\hbar \gamma^2}{l}}\left\{\sum_{j=2}^{b}\, \frac{1}{\sqrt{2\omega_j}}\left(
A_je^{-i\omega_j t}+A_j^*e^{i\omega_j t}\right)f_{\omega_j}(x)\right. \\ &+& \left.\sum_{n=-\infty}^\infty\, \frac{1}{\sqrt{2\omega(q_n)}}\left(
A(q_n)e^{-i\omega(q_n) t}f_{q_n}^*(x)+A(q_n)^*e^{i\omega(q_n) t}f_{q_n}(x)\right)\right\} \qquad .
\end{eqnarray*}
We have not included in this expansion the zero mode because this mode does not enter at one-loop order (its existence is due to the spontaneous translational symmetry breaking). Also, the possible half-bound state arises only at the $l=\infty$ limit when the energy of the highest bound state is exactly the energy of the scattering threshold. In that case, the Levinson theorem also sets a $\frac{1}{2}$-weight to this state; see \cite{Alonso2004}.
In quantum theory, the coefficients of the expansion become creation and annihilation operators of mesons moving either on the kink classical background -the scattering states-, or trapped by the kink: the bound states. The shift in the kink classical energy due to these unoccupied one-loop kink fluctuations is:
\[
\bigtriangleup\tilde{E}_0[\psi_K]=\frac{m_d}{\gamma_d^2}\bigtriangleup E_0[\phi_K]=\frac{\hbar m_d}{2}\left(\lim_{N\to\infty}\sum_{-N}^N\, \omega(q_n)+\sum_{j=2}^{b}\, \omega_j\right) \quad ,
\]
Sending the length $l$  of the normalization interval and the cutoff in the energy $N$ to infinity, this formula becomes:
\begin{eqnarray*}
\bigtriangleup E_0[\phi_K]&=&\frac{\hbar\gamma_d^2}{2}\left(\int_{-\infty}^\infty \, dk \, \rho(k)\omega(k)+\sum_{j=1}^{b-1}\, \omega_j+s_b\omega_b\right)\\ &=&\frac{\hbar\gamma_d^2}{2}\left(\int_{-\infty}^\infty \, dk \, \left(\frac{l}{2\pi}+\frac{1}{2\pi}\frac{\partial\delta}{\partial k}(k)\right)\sqrt{k^2+v^2}+\sum_{j=1}^{b-1}\, \omega_j+s_b\omega_b\right) \quad ,
\end{eqnarray*}
where $s_b=1$ if $\omega_b<v$ and $s_b=\frac{1}{2}$ if $\omega_b=v$.

\section{Kink mass quantum correction. The DHN formula}

It is clear that both $\bigtriangleup E[\phi^{(i)}]$ and $\bigtriangleup E[\phi_K]$ are divergent quantities (series
or integrals) because there is an infinite number of fluctuation modes. In the mid-seventies, Dashen, Hasslacher, and
Neveu implemented a regularization and renormalization procedure in two steps.
\subsection{Zero-point energy renormalization: mode by mode subtraction}

In the first step, a zero point energy renormalization is performed by means of a mode-by-mode substraction. This kink
mode-by-mode energy subtraction can be written symbolically as:
\begin{equation}
\bigtriangleup E_1[\phi_K]=\frac{\hbar\gamma_d^2}{2}{\rm Tr}_{L^2}\, \left(K^{\frac{1}{2}}-K_0^{\frac{1}{2}}\right)\label{kincas} =\frac{\hbar\gamma_d^2}{2}\lim_{N\to\infty}\sum_{1}^{N}\, \left(\omega_j-\nu_j\right)\nonumber\quad ,
\end{equation}
where the index $j=1,2, \cdots, N$ labels the eigenvalues ordered in increasing energy of both operators, when the spectra are discrete because some boundary conditions are chosen at the frontier of some finite interval of length $l$.
In the Appendix of \cite{Alonso2004}, however, we showed, using the one dimensional Levinson theorem contextualized in the kink and vacuum fluctuation spectra described above, that the subtle $l=\infty$ limit produces the formula:
\begin{equation}
\bigtriangleup E_1(\phi_K)=\frac{\hbar\gamma_d^2}{2}\left(\int_0^\infty \, \frac{dk}{\pi}\,\cdot \frac{\partial\delta}{\partial k}\cdot\sqrt{k^2+v^2}+\frac{1}{2\pi}\langle V(x) \rangle +\sum_{j=2}^{b-1}\, \omega_j+s_b\omega_b-\frac{v}{2}\right) \label{kcas} \quad ,
\end{equation}
where $\langle V(x) \rangle =\int_{-\infty}^\infty \, dx \, V(x)$. The two last terms in (\ref{kincas}) collect the contribution of the possible half-bound state of $K$ and the ever present half-bound state of $K_0$. In the sG and $\phi^4$ models, the net effect is null but we shall consider other kinks such that there are no half-bound states in the spectrum of $K$, thus generalizing the DHN formula to the form shown in (\ref{dhn}). We remark that, in our derivation of this formula, again following \cite{Barton}, we relied on even and odd states complying with Dirichlet boundary conditions. Although not usual in QFT, this approach is convenient for two reasons: 1) There is no degeneracy between eigenstates with opposite wave number, which helps to distinguish the half-bound state from the continuous threshold. 2) The even state of energy $v^2$ has wave vector $\frac{1}{2}$, which gives rise (rather indirectly) to the $\frac{1}{2}$-weight.

In the standard literature on the subject, the divergent integral is regularized by means of a cutoff:
\[
\bigtriangleup E_1(\phi_K)[\Lambda]=\frac{\hbar\gamma_d^2}{2} \left(\int_0^\Lambda \, \frac{dk}{\pi}\,\cdot \frac{\partial\delta}{\partial k}\cdot\sqrt{k^2+v^2}+\frac{1}{2\pi}\langle V(x) \rangle +\sum_{j=2}^{b-1}\, \omega_j+s_b\omega_b-\frac{v}{2}\right)
\]
such that $\bigtriangleup E_1(\phi_K)=\lim_{\Lambda\rightarrow \infty} \bigtriangleup E_1(\phi_K)[\Lambda]$. We say that the kink Casimir energy is regularized by means of a cutoff $\Lambda$ in the number of modes.

\subsection{Mass renormalization}

$\bigtriangleup E_1(\phi_K)$ in (\ref{kcas}) is still divergent because in models with interactions there are more than the vacuum energy ultraviolet divergences. In $(1+1)$-dimensional scalar field theory normal ordering tames all the uv divergences. The contraction of two field operators at the same point is equal to the normal ordered product plus a divergent term coming from the one-loop graph in Figure 1 below:
\[
\hat{\phi}^2(x^\mu)= :\hat{\phi}^2(x^\mu):+\delta v^2 \, \, \,\qquad  , \qquad \quad
\delta v^2 =\frac{1}{2l}\sum_{n\in{\mathbb Z}}\, \frac{1}{(\frac{4\pi^2}{l^2}n^2+v^2)^\frac{1}{2}}=\int \frac{dk}{4\pi}\, \frac{1}{\sqrt{k^2+v^2}} \quad ,
\]
where we have used periodic boundary conditions $f(-\frac{l}{2})=f(\frac{l}{2})$ -henceforth plane waves as one-particle states- to place the system in a normalization interval of non dimensional length $l$ as usual in QFT
{\footnote{Henceforth, we shall use PBC. At the $l=\infty$ limit, however, we shall
incorporate the contributions of the half-bound states derived in the ultra-long distance regime of the spectrum
with Dirichlet boundary conditions.}}.

\begin{figure}[ht]
\centerline{\includegraphics[height=1.5cm]{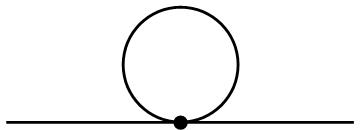}}
\caption{\textit{Self-energy graph due to the valence-four vertex with factor $\frac{\partial^2U}{\partial\phi^2}(\phi^{(i)})$.} }
\end{figure}
The Hamiltonian density operator
\[
\hat{\cal H}[\hat{\phi}(x^\mu)]=\frac{1}{2}\left(\hat{\pi}^2(x^\mu)+\frac{\partial\hat{\phi}}{\partial x} \frac{\partial\hat{\phi}}{\partial x}(x^\mu)\right)+U(\hat{\phi}(x^\mu))\quad ,
\]
normal ordered in the standard form (Wick theorem),
\[
:\hat{\cal H}:=\hat{\cal H}+:\left(
1-{\rm exp}\left[\frac{\hbar\gamma_d^2}{2}\delta v^2\frac{\delta^2}{\delta\phi^2}\right]\right)U(\hat{\phi}(x^\mu)):=\hat{\cal H}-\frac{\hbar\gamma_d^2}{2}\delta v^2:\frac{\partial^2U}{\partial\phi^2}(\hat{\phi}(x^\mu)):+{\cal O}\left(\hbar^2\gamma_d^4\delta v^4:\frac{\partial^4U}{\partial\phi^4}(\hat{\phi}):\right)\quad ,
\]
shows the necessary one-loop counter term{\footnote{In 2D scalar field theory there are only ultraviolet divergences in closed-loop diagrams with only one internal line. Normal ordering is enough to control ultraviolet divergences.}}. Therefore, the self-energy counter-term contributes to the one-loop kink mass shift:
\[
\bigtriangleup E_2(\phi_K)= -\frac{\hbar\gamma_d^2}{2}\delta v^2\int_{-\infty}^\infty \, dx \, \left(\langle 0;K|:\frac{\partial^2U}{\partial\phi^2}(\hat{\phi}):|K;0\rangle-
\langle 0;V|:\frac{\partial^2U}{\partial\phi^2}(\hat{\phi}):|V;0\rangle\right)\quad .
\]
Here, $|0;K\rangle$ and $|0;V\rangle$ are respectively the ground states - no mesons at all - in the kink and
constant vacuum backgrounds. These states are coherent states and the action of normal ordered operators in this type of quantum states is easily calculated:
 \[
 \langle 0;K|:\frac{\partial^2U}{\partial\phi^2}(\hat{\phi}):|K;0\rangle-
\langle 0;V|:\frac{\partial^2U}{\partial\phi^2}(\hat{\phi}):|V;0\rangle=\frac{\partial^2U}{\partial\phi^2}(\phi_K(x))-
 \frac{\partial^2U}{\partial\phi^2}(\phi^{(i)})=V(x) \quad .
 \]
Consequently,
\[
\bigtriangleup E_2(\phi_K) = -\frac{\hbar\gamma_d^2}{8\pi}  \left<V(x) \right>  \int_{-\infty}^\infty \frac{dk}{\sqrt{k^2+v^2}} \quad ,
\]
and the DHN formula for the one-loop kink mass shift reads:
\begin{eqnarray}
\frac{\bigtriangleup E(\phi_K)}{\hbar\gamma_d^2}&=&\frac{\bigtriangleup E_1(\phi_K)+\bigtriangleup E_2(\phi_K)}{\hbar\gamma_d^2}
 =\frac{1}{2}\sum_{j=2}^{b-1} \omega_j + \frac{1}{2} s_b \omega_b - \frac{v}{4}+ \label{dhn}\\ &+& \frac{1}{2\pi} \int_0^\infty dk \frac{\partial \delta(k)}{\partial k} \sqrt{k^2 + v^2} + \frac{1}{4\pi} \left<V(x) \right>\left(1 -\frac{1}{2}\int_{-\infty}^\infty \frac{dk}{\sqrt{k^2+v^2}}\right)\quad  ,
\nonumber
\end{eqnarray}
which has been obtained from the regularized version when the cutoff goes to infinity:
\begin{eqnarray}
\frac{\bigtriangleup E(\phi_K)}{\hbar\gamma_d^2}&=&\lim_{\Lambda\to\infty}\frac{\bigtriangleup E(\phi_K)[\Lambda]}{\hbar\gamma_d^2}
 =\frac{1}{2}\sum_{j=2}^{b-1} \omega_j + \frac{1}{2} s_b \omega_b - \frac{v}{4}+ \label{dhncr}\\ &+& \frac{1}{2\pi} \lim_{\Lambda\to\infty}\int_0^\Lambda dk \frac{\partial \delta(k)}{\partial k} \sqrt{k^2 + v^2} + \frac{1}{4\pi} \left<V(x) \right>\left(1 -\frac{1}{2}\lim_{\Lambda\to\infty}\int_{-\Lambda}^\Lambda\frac{dk}{\sqrt{k^2+v^2}}\right)\nonumber\quad  .
\end{eqnarray}

\subsubsection{The prototypes: the sine-Gordon and $\lambda\phi^4$ kinks.}

In the sine-Gordon model we have:
\begin{eqnarray*}
K_0 &=& -\frac{d^2}{dx^2}+1 \hspace{0.3cm}, \hspace{0.3cm} K = -\frac{d^2}{dx^2}+1 - 2 \,{\rm sech}^2 x \hspace{0.3cm}, \hspace{0.3cm} \langle V(x)\rangle =-4 \\ {\rm Spec}\,K_0 &=&\{\nu_1=1\}_\frac{1}{2} \cup \{\nu^2(k)=k^2+1\}_{k\in \mathbb{R}} \\ {\rm Spec}\,K &=&\{\omega_1=0\} \cup \{\omega_2=1\}_\frac{1}{2} \cup \{\omega^2(q)=q^2+1\}_{q\in \mathbb{R}} \\ \rho_0(k)&=&\frac{l}{2\pi} \quad , \quad \delta(q)=2{\rm arctan}\frac{1}{q} \quad , \quad
\rho(q)=\frac{l}{2\pi}+ \frac{1}{2\pi} \frac{d\delta(q)}{dq}=\frac{l}{2\pi}-\frac{1}{\pi}\frac{1}{1+q^2}
\end{eqnarray*}
Discarding the zero mode in the spectrum of $K$ the eigenvalues of $K_0$ and $K$ are the same: one half-bound state and a continuous spectrum starting at the threshold $\omega^2=1$. Because of the phase shifts induced by the kink potential well the spectral densities are different. A mode-number cutoff regulator applied to the DHN formula provides the exact result:
\[ \frac{\bigtriangleup E(\phi_K)}{\hbar\gamma_d^2}
=-\frac{1}{\pi}\lim_{\Lambda\to\infty}\int_0^\Lambda\, dk \, \frac{1}{\sqrt{1+k^2}}-\frac{1}{\pi}\left(1-\lim_{\Lambda\to\infty}\int_0^\Lambda\, dk \, \frac{1}{\sqrt{k^2+1}}\right)=-\frac{1}{\pi} \quad .
\]
Note that a cutoff in the energy would leave $\frac{\bigtriangleup E(\phi_K)}{\hbar\gamma_d^2}=0$ as the one-loop kink mass shift.

For $\lambda\phi^4$ kinks things are sligthly more involved:
\begin{eqnarray*}
K_0 &=& -\frac{d^2}{dx^2}+4 \hspace{0.3cm}, \hspace{0.3cm} K = -\frac{d^2}{dx^2}+4 - 6 \,{\rm sech}^2 x \hspace{0.3cm}, \hspace{0.3cm} \langle V(x)\rangle =-12 \\ {\rm Spec}\,K_0 &=&\{\nu_1=4\}_\frac{1}{2} \cup \{\nu^2(k)=k^2+4\}_{k\in \mathbb{R}} \\ {\rm Spec}\,K &=&\{\omega_1=0\} \cup \{\omega_2^2=3\} \cup \{\omega_3^2=4\}_\frac{1}{2} \cup \{\omega^2(q)=q^2+4\}_{q\in \mathbb{R}} \\ \rho_0(k)&=&\frac{l}{2\pi} \quad , \quad \delta(q)=-2{\rm arctan}\frac{3q}{2-q^2 } \quad , \quad
\rho(q)=\frac{l}{2\pi}+ \frac{1}{2\pi} \frac{d\delta(q)}{dq}=\frac{l}{2\pi}-\frac{3}{\pi}\frac{2+q^2}{(1+q^2)(4+q^2)}
\end{eqnarray*}
In this system, besides the zero mode there is another bound state in the spectrum of $K$ with eigenvalue: $\omega_2^2=3$. All the other eigenvalues of $K_0$ and $K$ are the same: one half-bound state, $\omega_3^2=4$, and a continuous spectrum starting at the threshold:
 $\omega^2(q)=q^2+v^2$. The phase shifts induced by the kink potential well cause different spectral densities as in the sine-Gordon model. The mode-number cutoff regulator applied in the DHN formula also provides the exact result:
\begin{eqnarray*} \frac{\bigtriangleup E(\phi_K)}{\hbar\gamma_d^2}
&=&\frac{\sqrt{3}}{2}-\frac{3}{\pi}\lim_{\Lambda\to\infty}\int_0^\Lambda\, dk \, \frac{2+k^2}{\sqrt{(1+k^2)(4+k^2)}}-\frac{3}{\pi}\left(1-\lim_{\Lambda\to\infty}\int_0^\Lambda\, dk \, \frac{1}{\sqrt{k^2+4}}\right)\\&=& \frac{\sqrt{3}}{2}-\frac{3}{\pi}\lim_{\Lambda\to\infty}\left({\rm arcsinh}(\frac{\Lambda}{2})+\frac{1}{\sqrt{3}}{\rm arctan}(\frac{\sqrt{3}\Lambda}{\sqrt{4+\Lambda^2}})\right)-\frac{3}{\pi}\left(1-\lim_{\Lambda\to\infty}{\rm arcsinh}(\frac{\Lambda}{2})\right)\\&=&\frac{\sqrt{3}}{2}- \frac{3}{\pi} -\frac{1}{\sqrt{3}}=\frac{1}{2\sqrt{3}}-\frac{3}{\pi}\quad .
\end{eqnarray*}
A cutoff in the energy provides the answer: $\frac{\bigtriangleup E(\phi_K)}{\hbar\gamma_d^2}=\frac{1}{2\sqrt{3}}$.

\section{Zeta function regularization procedure}

The problem with the DHN formula is that one needs complete spectral information about the $K$
operators governing the kink fluctuations and, at present, it is available only for the sine-Gordon and $\lambda(\phi)^4_2$ kinks. Instead, we shall rely on the spectral $K$-heat kernel expansion for computing
one-loop kink mass shifts in other interesting models where there is not enough spectral information about the kink fluctuations. However, before this we shall regularize the DHN formula by the zeta function procedure to test the conceptual nature of this method on sine-Gordon and $\lambda\phi^4$ kinks.

The traces and determinants of powers of elliptic operators are only
defined by means of a process of analytic continuation that mimics
the definition of the Riemann zeta function as a meromorphic
function, provideng formal meaning to strictly divergent series in some
regions (${\rm Re}\, s<\frac{1}{2}$) of the complex plane. By replacing
natural numbers by eigenvalues (hopefully forming a discrete
spectrum), generalized zeta functions associated with differential
operators can be defined. If we assume that $A\in {\cal A}[\bar{L}^2(\mathbb{R})]$ is an operator with a positive discrete spectrum in the form of ${\rm Spec}\,A=\{\omega_n^2 \in \mathbb{R}:A\psi_n=\omega_n^2 \psi_n\}$, the generalized zeta function associated with the operator $A$ is a meromorphic function in the complex plane defined as
\[
\zeta_A(s)={\rm Tr}\, A^{-s} = \sum_n \frac{1}{(\omega_n^2)^s} \hspace{0.5cm}, \hspace{0.5cm} s\in \mathbb{C}
\]
There is a general theory of elliptic
pseudo-differential operators that characterizes the conditions
under which the generalized zeta functions are meromorphic
functions, and the values apart from the poles of the zeta function and
the derivatives of zeta, are taken as \lq\lq regularized" definitions of
traces and logarithms of determinants of (complex powers of
pseudo-)differential operators \cite{Gilkey1984, Roe1988, Elizalde1994, Kirsten2002, Vassilevich2003}.

Direct computation of the generalized zeta function is difficult in general. In order to identify the poles of $\zeta_{A}(s)$, we introduce a non-dimensional fictitious temperature $\beta=\frac{\hbar m_d}{k_B T}$ {\footnote{Note that, in our system of units, the Boltzmann constant $k_B$ and  fictitious temperature $T$ respectively have dimensions of $ML$ and $L^{-1}$.}} and define the $A$-heat function:
\begin{equation}
h_A(\beta)={\rm Tr}\, e^{-\beta A} = \sum_{n=0} e^{-\beta \omega_n^2}
\label{heatfunction}
\end{equation}
which has better convergence properties. Note that for operators with a positive spectrum the low-temperature asymptotic condition
\[
\lim_{\beta \rightarrow \infty} h_A(\beta)=0
\]
holds. The $\zeta_{A}(s)$ and the $h_A(\beta)$ functions are related through the Mellin transform:
\begin{equation}
\zeta_A(s)= \frac{1}{\Gamma(s)} \int_0^\infty d\beta \beta^{s-1} h_A(\beta) \quad .
\label{Mellin}
\end{equation}

\subsection{Vacuum spectral zeta and heat functions}

We shall now consider the $L^2$ trace of the $K_0$ operator using periodic boundary conditions on the normalization interval. In the previous Section, we mentioned that Dirichlet boundary conditions are convenient to disentangle the half-bound state that arises at the $l=\infty$ limit from the threshold of the continuous spectrum through the one-dimensional Levinson theorem. We now know how to correct the kink Casimir energy to take this subtle effect into account.

The energy of the free Hamiltonian evaluated at the vacuum state is:
\[
\bigtriangleup\tilde{E}(\psi^{(i)})=\langle 0;V|\hat{H}^{(2)}|V;0\rangle=\frac{\hbar m_d}{2}{\rm Tr}_{L^2({\mathbb S}^1)}\, K_0^{\frac{1}{2}}=\frac{\hbar m_d}{2}\sum_{n\in{\mathbb Z}}\, \left[\frac{4\pi^2}{l^2}n^2+v^2\right]^{\frac{1}{2}}=\frac{\hbar m_d}{2}\zeta_{K_0}(-\frac{1}{2}) \quad .
\]
Here, $\hat{H}^2$ is the free Hamiltonian; i.e., the quadratic Hamiltonian in the fluctuations around the vacuum,
$|V;0\rangle$ is the vacuum state where all the fluctuation states are unoccupied, and $\zeta_{K_0}(-\frac{1}{2})$ is
the generalized/spectral zeta function of the differential operator $K_0$ at the $s=-\frac{1}{2}$ point in the $s$-complex plane. In fact, one considers in general the trace of $K_0^{-s}$
\[
\zeta_{K_0}(s)={\rm Tr}_{L^2({\mathbb S}^1)}\, K_0^{-s}=\sum_{n\in{\mathbb Z}}\, \frac{1}{\left[\frac{4\pi^2}{l^2}n^2+v^2\right]^s}=
E[s,v^2|{\textstyle\frac{4\pi^2}{l^2}}] \, \, , \quad s\in{\mathbb C} \quad ,
\]
which is a convergent series only for ${\rm Re}(s)>\frac{1}{2}$ but can be analytically continued to the whole
$s$-complex plane to find the meromorphic Epstein function, $E[s,v^2|\frac{4\pi^2}{l^2}]$. We also define the heat function:
\[
h_{K_0}(\beta)={\rm Tr}_{L^2({\mathbb S}^1)}e^{-\beta K_0}=e^{-v^2\beta}\sum_{n\in{\mathbb Z}}\,e^{-\beta\frac{4\pi^2}{l^2}n^2} =e^{-v^2\beta}\Theta\left[\begin{array}{c} 0\\
0\end{array}\right]\left(0|i {\textstyle\frac{4\pi}{l^2}}\beta\right)
\]
where $\Theta\left[\begin{array}{c} a \\
b\end{array}\right](z|\tau)=\sum_{n\in{\mathbb Z}}e^{2\pi
i[(n+a)(z+b)+\frac{(n+a)^2}{2}\tau)]}$ ($a,b=0,\frac{1}{2}$) in the form of a series that is a Riemman  Theta constant: a Riemman Theta function at $z=0$. The spectral $K_0$-zeta and $K_0$-heat functions are related via the Mellin transform:
 \[
 \zeta_{K_0}(s)=\frac{1}{\Gamma(s)}\int_0^\infty\, d\beta \, \beta^{s-1}\, h_{K_0}(\beta) \, \Leftrightarrow \, E[s,v^2|{\textstyle\frac{4\pi^2}{l^2}}]=\frac{1}{\Gamma(s)}\int_0^\infty\, d\beta \, \beta^{s-1}\, e^{-v^2\beta}\Theta\left[\begin{array}{c} 0\\
0\end{array}\right] (0|i{\textstyle\frac{4\pi}{l^2}}\beta) \quad .
\]
The modular transformation $\tau=i\frac{4\pi}{l^2}\beta \, \rightarrow \, -\frac{1}{\tau}=i\frac{l^2}{4\pi\beta}$, or alternatively the Poisson summation formula, connects our Riemann Theta constant, which is a modular function of weight $-\frac{1}{2}$, to another Theta constant in the form:
\[
\Theta\left[\begin{array}{c} 0\\
0\end{array}\right](0|i {\textstyle \frac{4\pi}{l^2}} \beta)=\frac{l}{\sqrt{4\pi\beta}}\Theta\left[\begin{array}{c} 0\\
0\end{array}\right](0|i {\textstyle\frac{l^2}{4\pi\beta}}) \quad .
\]
The Mellin transform of this Poisson-inverted formula gives
\begin{eqnarray}
\zeta_{K_0}(s)&=&\frac{l}{\sqrt{4\pi}\Gamma(s)}\int_0^\infty \, d\beta \, \beta^{s-\frac{3}{2}}\sum_{n=-\infty}^\infty \, e^{-v^2\beta-\frac{n^2l^2}{4\beta}} \label{zetav1}\\ &=& \frac{l}{\sqrt{4\pi}\Gamma(s)}\left[\frac{\Gamma(s-\frac{1}{2})}{v^{2s-1}}+2\sum_{n\in{\mathbb Z}/\{0\}}\, \left(\frac{nl}{2}\right)^{s-\frac{1}{2}}K_{\frac{1}{2}-s}(nl)\right] \label{zetav2} \quad .
\end{eqnarray}
In formula (\ref{zetav2}), the Kelvin functions $K_{\frac{1}{2}-s}(ln)$ are transcendental entire functions of $s$ (holomorphic for every $s\in{\mathbb C}/\infty$, with an essential singularity at $s=\infty$). Therefore, the poles
$\zeta_{K_0}(s)$ are the  poles of the Euler $\Gamma(s-\frac{1}{2})$ function:
\[
s=\frac{1}{2}, -\frac{1}{2}, -\frac{3}{2}, -\frac{5}{2}, \cdots , -\frac{2j+1}{2} , \, \, \cdots \, \, , \, \, j\in{\mathbb Z}^+ \quad .
\]
The lengthy asymptotics, where the Hilbert spaces on the circle and the real line are traded, $L^2({\mathbb S}^1)\to \bar{L}^2({\mathbb R})$, is easily read from formula (\ref{zetav1}). The Mellin transform is then applied to obtain:
\begin{equation}
 h_{K_0}(\beta)=\frac{l}{2\pi}\int_{-\infty}^\infty \, dk\, e^{-\beta(k^2+v^2)}=\frac{le^{-v^2\beta}}{\sqrt{
4\pi\beta}} \quad , \quad \zeta_{K_0}(s)=\frac{l}{2\pi}\int_{-\infty}^\infty\frac{dk}{(k^2+v^2)^s}=\frac{l}{\sqrt{4\pi}v^{2s-1}}
\frac{\Gamma(s-\frac{1}{2})}
{\Gamma(s)} \quad .
\end{equation}
For later use it is also interesting to consider the
high-temperature, small $\beta$, asymptotics:
\[
\Theta\left[\begin{array}{c} 0 \\
0\end{array}\right](0|i {\textstyle\frac{l^2}{4\pi\beta}})=\sum_{n\in{\mathbb
Z}}\, e^{-\frac{l^2}{4\beta}n^2}\cong_{\beta\to 0}\, 1+{\cal
O}(e^{-\frac{c}{\beta}}) \quad .
\]
The idea of zeta function regularization now follows naturally: simply assign to the infinite vacuum energy a finite
value provided by the spectral zeta function at a regular point in $s\in{\mathbb C}$:
\begin{equation}
\bigtriangleup E(\phi^{(i)})=\frac{\hbar \gamma_d^2}{2}\zeta_{K_0}(-{\textstyle\frac{1}{2}}) \, \, \rightarrow \, \, \bigtriangleup E(\phi^{(i)})[s]=\frac{\hbar \gamma_d^2}{2}\frac{\mu}{m_d}\left(\frac{\mu^2}{m_d^2}\right)^s\zeta_{K_0}(s) \quad ,
\end{equation}
where $\mu$ is a parameter of dimensions $L^{-1}$ introduced to keep the dimensions of the regularized energy right.
We stress that $s=-\frac{1}{2}$ is a pole of this function.

\subsection{Kink spectral zeta and heat functions}

The spectral $K$-heat function with periodic boundary conditions at the finite length interval is:
\begin{equation}
h_K(\beta)={\rm Tr}_{L^2({\mathbb S}^1)}e^{\beta K}=1+\sum_{j=2}^{b}\, e^{-\beta \omega_j^2} + e^{-\beta v^2}\sum_{q_n\in {\rm Ker}\,\Phi (q,l,n)} \, e^{-\beta q_n^2} \label{khfs}
\end{equation}
where the $q_n$ satisfy the transcendental spectral conditions
\[
\Phi(q,l,n)=q+\frac{1}{l}\delta(q)-\frac{2\pi}{l}n=0 \qquad .
\]
There is now a delicate point to be mentioned before of performing the Mellin transform. The zero mode $\omega_1^2=0$ in the spectrum of $K$ introduces a spurious contribution in the spectral $K$-zeta function. In fact
\begin{equation}
\lim_{\varepsilon\to 0}\frac{1}{\Gamma(s)}\int_0^\infty \, d\beta \, \beta^{s-1}\, e^{-\varepsilon\beta} =\lim_{\varepsilon\to 0}\frac{1}{\varepsilon^s} \label{zzm}
\end{equation}
valid only if ${\rm Re}(s)>1$, and it diverges in the $\varepsilon\to 0$ limit when it is defined.
It is known, however, that the zero mode does not contribute to the one-loop quantum correction $\bigtriangleup E(\phi_K)$ \cite{Rajaraman1982} but only enters to second order in the $\hbar$ expansion. Fortunately, within the
zeta function philosophy, the result (\ref{zzm}) can be extended to the whole $s$-complex plane and it is 0 if ${\rm Re}(s)<0$. Thus, we shall perform the Mellin transform of the $K$-heat function on the subspace orthogonal to the
kernel of the $K$ operator without any loss of information regarding our problem:
\begin{eqnarray}
&& {\rm Tr}_{L^2({\mathbb S}^1)}e^{-\beta K^\bot}={\rm Tr}_{L^2({\mathbb S}^1)}e^{-\beta K}-1 \quad , \quad \zeta_{K^\bot}(s)=\frac{1}{\Gamma(s)}\int_0^\infty\, d\beta \, \beta^{s-1}{\rm Tr}_{L^2({\mathbb S}^1)}e^{-\beta K^\bot} \nonumber \\ && \zeta_{K^\bot}(s)=\sum_{j=2}^{b}\, \frac{1}{\omega_j^{2s}}+\sum_{q_n\in {\rm Ker}\,\Phi (q,l,n)} \, \frac{1}{(q_n^2+v^2)^s} \quad . \label{kzfb}
\end{eqnarray}
At the $l\to\infty$ limit, the regularized $K$-heat and $K$-zeta functions become:
\begin{eqnarray}
&& h_{K^\bot}(\beta)={\rm Tr}_{\bar{L}^2({\mathbb R})}e^{-\beta K^\bot}=\sum_{j=2}^{b-1}\, e^{-\beta\omega_j^2} + e^{-\beta(s_b^2\omega_b^2)}+ \int_{-\infty}^\infty \, \frac{dk}{2\pi}\left(l+\frac{\partial\delta}{\partial k}\right)e^{-\beta (k^2+v^2)}
\label{khfsl} \\ && \zeta_{K^\bot}(s)={\rm Tr}_{\bar{L}^2({\mathbb R})}\left(K^\bot\right)^{-s}=\sum_{j=2}^{b-1}\, \frac{1}{\omega_j^{2s}}+\frac{1}{s_b^{2s}\omega_b^{2s}}+\, \int_{-\infty}^\infty \, \frac{dk}{2\pi}\left(l+\frac{\partial\delta}{\partial k}\right)\frac{1}{(k^2+v^2)^s} \quad . \label{kzfbl}
\end{eqnarray}
Thus, the zeta function regularized kink Casimir energy reads as follows:
\begin{eqnarray}
\bigtriangleup E_1(\phi_K)[s]&=& \bigtriangleup E(\phi_K)[s]-\bigtriangleup E(\phi^{(i)})[s]=\frac{\hbar \gamma_d^2}{2}\left(\frac{\mu^2}{m_d^2}\right)^{s+\frac{1}{2}}\left(\zeta_{K^\bot}(s)-\zeta_{K_0}(s)\right)\nonumber
\\ &=&\sum_{j=2}^{b-1}\, \frac{1}{\omega_j^{2s}}+\frac{1}{s_b^{2s}\omega_b^{2s}}-\frac{4^{2s}}{v^{2s}} +\, \int_{-\infty}^\infty \, \frac{dk}{2\pi}\frac{\partial\delta}{\partial k}\frac{1}{(k^2+v^2)^s} \label{zrkc}\quad .
\end{eqnarray}

The energy due to the one-loop mass counter-term
\[
\bigtriangleup E_2(\phi_K)=-\frac{\hbar \gamma_d^2}{8\pi}\langle V(x) \rangle \int_{-\infty}^\infty \frac{dk}{\sqrt{k^2+v^2}}=- \frac{\hbar \gamma_d^2}{4}\langle V(x) \rangle \lim_{l\to\infty}\frac{1}{l}
\sum_{n\in{\mathbb Z}}\, \frac{1}{\sqrt{\frac{4\pi^2}{l^2}n^2+v^2}}
\]
is also regularized by the zeta function procedure:
\[
\bigtriangleup E_2(\phi_K)[s]=  \frac{\hbar \gamma_d^2}{2}\langle V(x) \rangle \left(\frac{\mu^2}{m_d^2}\right)^{s+\frac{1}{2}}\lim_{l\to\infty}\frac{1}{l}\frac{\Gamma(s+1)}{\Gamma(s)}\zeta_{K_0}(s+1)
\quad .
\]
Note that we regularize this quantity by means of $\zeta_{K_0}(s+1)$. This is necessary to compare residua at the same point of the $s$-complex plane. Finally, we obtain the zeta function regularized DHN formula:
\begin{eqnarray}
\hspace{-0.2cm}&&\bigtriangleup E(\phi_K)=\nonumber \\ \hspace{-0.2cm}&=&\frac{\hbar \gamma_d^2}{2}\lim_{s\to -\frac{1}{2}}\left(\frac{\mu^2}{m_d^2}\right)^{s+\frac{1}{2}}\frac{1}{\Gamma(s)}\left\{\int_0^\infty \, d\beta \, \beta^{s-1}\left(h_{K^\bot}(\beta)-h_{K_0}(\beta)\right)+\frac{\langle V(x) \rangle}{2}\lim_{l\to\infty}\frac{\Gamma(s+1)}{l}\zeta_{K_0}(s+1)\right\} \label{zolqc}  \\ \hspace{-0.2cm}&=& \frac{\hbar \gamma_d^2}{2}\lim_{s\to -\frac{1}{2}}\left(\frac{\mu^2}{m_d^2}\right)^{s+\frac{1}{2}}\frac{1}{\Gamma(s)}\left\{\sum_{j=2}^{b-1}\, \frac{1}{\omega_j^{2s}}+\frac{1}{s_b^{2s}\omega_b^{2s}}-\frac{4^{2s}}{v^{2s}} +\, \int_{-\infty}^\infty \, \frac{dk}{2\pi}\frac{\partial\delta}{\partial k}\frac{1}{(k^2+v^2)^s} +\frac{\langle V(x) \rangle}{2}\frac{\Gamma(s+\frac{1}{2})}{v^{2s+1}}\right\}\nonumber\quad .
\end{eqnarray}

\subsubsection{The prototypes: sine-Gordon and $\lambda\phi^4$ kinks}

\begin{itemize}

\item In the sine-Gordon model we have:
\begin{eqnarray*}
h_{K_0}(\beta)&=& \frac{l}{\sqrt{4\pi\beta}}e^{-\beta} \qquad , \qquad \zeta_{K_0}(s)=
\frac{l}{\sqrt{4\pi}}\frac{\Gamma(s-\frac{1}{2})}{\Gamma(s)}\\ h_{K^\bot}(\beta)&=& h_{K_0}(\beta)-{\rm Erfc}(\sqrt{\beta}) \qquad , \qquad \zeta_{K^\bot}(s)=\zeta_{K_0}(s)-\frac{1}{\sqrt{\pi}}\frac{\Gamma(s+\frac{1}{2})}{s\Gamma(s)}\quad .
\end{eqnarray*}
We can now compute the limit of the regularized quantities that
enter the one-loop correction formula to the kink mass:
\begin{eqnarray*}
\Delta E_1(\phi_K)&=&\frac{\hbar\lambda}{2m^2} \left[
\lim_{s\rightarrow -\frac{1}{2}} \left( \frac{\mu^2}{m^2}
\right)^{s+\frac{1}{2}} (\zeta_{K^\bot}(s)-\zeta_{K_0}(s))\right] \\
&=& -\frac{\hbar\lambda}{2m^2} \left[
\lim_{s\rightarrow -\frac{1}{2}} \left( \frac{\mu^2}{m^2}
\right)^{s+\frac{1}{2}} \frac{\Gamma(s+\frac{1}{2})}{\sqrt{\pi}s\Gamma(s)}\right] \quad .
\end{eqnarray*}
In order to fall in the pole carefully, we look at the behaviour of this function in its neighborhood
$s=-\frac{1}{2}+\varepsilon$, and we take the $\varepsilon\to 0$ limit:
\begin{eqnarray}
\Delta E_1(\phi_K)&=&-\frac{\hbar\lambda}{2\sqrt{\pi}m^2} \left[ \lim_{\varepsilon
\rightarrow 0} \left( \frac{\mu}{m} \right)^{2\varepsilon}
\frac{\Gamma (\varepsilon )}{\Gamma
(\frac{1}{2}+\varepsilon  )}\right ]\nonumber\\&=& \frac{\hbar
\lambda}{2\pi m^2}\lim_{\varepsilon\rightarrow 0}\left[
-\frac{1}{\varepsilon}-2\log\frac{\mu}{m}-\psi(1)+\psi(\frac{1}{2})+{\cal O}(\varepsilon
)\right ]\label{kcasreg} \quad ,
\end{eqnarray}
where $\psi(z)=\frac{\Gamma^\prime(z)}{\Gamma(z)}$ is the digamma function. Simili modo,
\begin{eqnarray*}
\Delta E_2(\phi^K) &=& -\frac{\hbar\lambda}{m^2}
\lim_{s\rightarrow -\frac{1}{2}} \left( \frac{\mu}{m}
\right)^{2s+1}\frac{\Gamma (s+1)}{\Gamma (s)} \lim_{l\to\infty}\frac{2}{l}\zeta_{K_0}(s+1) + {\cal O}(\hbar^2 \frac{\lambda^2}{m^4})\\ &=&-\frac{\hbar \lambda}{\sqrt{\pi}m^2}\lim_{s\to
-\frac{1}{2}}\left(\frac{\mu}{m}\right)^{2s+1}\frac{\Gamma
(s+\frac{1}{2})}{\Gamma (s)}+{\cal O}(\hbar^2 \frac{\lambda^2}{m^4})
\end{eqnarray*}
and
\begin{eqnarray}
\Delta E_2(\phi^K)&=&-\frac{\hbar
\lambda}{\sqrt{\pi}m^2}\left[\lim_{\varepsilon\to
0}\left(\frac{\mu}{m}\right)^{2\varepsilon}\frac{\Gamma
(\varepsilon)}{\Gamma (-\frac{1}{2}+\varepsilon)}\right] +{\cal O}(\hbar^2 \frac{\lambda^2}{m^4})\nonumber \\ &=& \frac{\hbar \lambda}{2 \pi m^2}
\lim_{\varepsilon \rightarrow 0} \left [
\frac{1}{\varepsilon}+2\log\frac{\mu}{m}+\psi(1)-\psi(-\frac{1}{2})+{\cal O}(\varepsilon)\right]+{\cal O}(\hbar^2 \frac{\lambda^2}{m^4})  \label{krenreg}\, \, \, .
\end{eqnarray}
The important point is that the renormalization of the
zero-point energy performed by the subtraction of $\zeta_{K_0}(-\frac{1}{2})$ still leaves a divergence coming from the $s=-\frac{1}{2}$ poles because the residua are different. The
correction due to the mass renormalization counter-term also has a
pole. The sum of these two contributions leaves a
finite remainder and we end with the finite answer:
\begin{equation}
\Delta E (\phi_K)= \Delta E_1(\phi_K) +\Delta E_2 (\phi_K)=-\frac{\hbar\lambda}{\pi m^2} \label{sgolms}\quad ,
\end{equation}
which is the same result as that obtained by cutting the number of modes in  the DHN formula. In sum, the one-loop renormalized kink mass can be written in the form:
\begin{eqnarray*}
\tilde{E}(\psi_{\rm 1-loop})&=&\tilde{E}(\psi_{\rm K})+ \Delta\tilde{E}(\psi_K)\\ &=& \frac{8 m^3}{\lambda}-\frac{\hbar
m}{\pi}+{\cal O}(\hbar^2\frac{\lambda^2}{m^4})=\frac{8 m^3}{\lambda}\left(1-\frac{\hbar}{8\pi}\frac{\lambda}{m^2}+{\cal O}(\hbar^2\frac{\lambda^2}{m^4})\right)=\frac{8 M^3}{\lambda} \quad .
\end{eqnarray*}
Note that if $\hbar\frac{\lambda}{m^2}>8\pi$, the coupling constant is greater than the famous bound \cite{Coleman1985} where the quantum sine-Gordon model does not exist. We find that, beyond Coleman's bound, $\tilde{E}(\psi_{\rm 1-loop})<0$ and the kink energy  would be lower than the vacuum energy. The renormalized particle mass $M$ is also pathological and becomes negative when $M^3<0$.

\item In the $\lambda\phi^4$ model, we have
\begin{eqnarray*}
h_{K_0}(\beta) &=& \frac{l}{2 \sqrt{2} \pi}
\int_{-\infty}^\infty \hspace{-0.2cm} dk\,
e^{-\beta(k^2+4)}=\frac{l}{\sqrt{8\pi \beta}}\, e^{-4 \beta} \quad , \quad
\zeta_{K_0}(s) = \frac{l}{\sqrt{8 \pi}} \frac{1}{4^{s-\frac{1}{2}}}
\frac{\Gamma(s-\frac{1}{2})}{\Gamma(s)}  \\
h_{K^\bot}&=& \frac{l}{\sqrt{8\pi \beta}}\, e^{-4 \beta} +e^{-3 \beta}
{\rm Erf}\sqrt{\beta}-{\rm Erfc}\, 2\sqrt{\beta}\, \\
\zeta_{K^\bot}(s)&=&\zeta_{K_0}(s)+
\frac{\Gamma(s+\frac{1}{2})}{\sqrt{\pi}\,\Gamma(s)} \left[
\frac{2}{3^{s+\frac{1}{2}}} {}_2F_{1}
[{\textstyle\frac{1}{2}},{s+\textstyle\frac{1}{2}},{\textstyle\frac{3}{2}},
-{\textstyle\frac{1}{3}}]-\frac{1}{4^s} \frac{1}{s} \right] \quad ,
\end{eqnarray*}
where ${}_2 F_1[a,b,c,d]$ is the Gauss hypergeometric function. The power expansion of ${}_2 F_1$,
\[
{}_2
F_{1}[{\textstyle\frac{1}{2}},{s+\textstyle\frac{1}{2}},{\textstyle\frac{3}{2}},
-{\textstyle\frac{1}{3}}]=\frac{\Gamma (\frac{3}{2})}{\Gamma
(\frac{1}{2})\Gamma (s+\frac{1}{2})}\sum_{j=0}^\infty
\frac{(-1)^j}{3^l j!}\frac{\Gamma (j+\frac{1}{2})\Gamma
(s+j+\frac{1}{2})}{\Gamma (j+\frac{3}{2})}
\]
tells us that, besides the poles of $\zeta_{K_0}(s)$,
$\zeta_{K^\bot}(s)$ has poles at
$s=-\frac{1}{2}+j,-\frac{3}{2}+j,-\frac{5}{2}+j,\cdots$, $j\in
{\mathbb Z}^{+} \cup \{0\}$; i.e., as in the sG soliton case,
$\zeta_{K_0}(s)$ and ${\zeta_{K^\bot}}(s)$ share
the same poles except $s=\frac{1}{2}$ but the residua in the
$\lambda (\phi^4)_2$ model are increasingly different with increasing values of $|{\rm Re}\hspace{0.1cm}{
 s}|$.

The kink Casimir energy is:
\begin{eqnarray*}
\Delta E_1(\phi_K)&=&\frac{\hbar\lambda}{2m^2} \left[
\lim_{s\rightarrow -\frac{1}{2}} \left( \frac{\mu^2}{m^2}
\right)^{s+\frac{1}{2}} (\zeta_{K^\bot}(s)-\zeta_{K_0}(s))\right] \\
&=& \frac{\hbar\lambda}{2m^2} \left[
\lim_{s\rightarrow -\frac{1}{2}} \left( \frac{\mu^2}{m^2}
\right)^{s+\frac{1}{2}}\frac{\Gamma(s+\frac{1}{2})}{\sqrt{\pi}\,\Gamma(s)} \left(
\frac{2}{3^{s+\frac{1}{2}}} {}_2F_{1}
[{\textstyle\frac{1}{2}},{s+\textstyle\frac{1}{2}},{\textstyle\frac{3}{2}},
-{\textstyle\frac{1}{3}}]-\frac{1}{4^s} \frac{1}{s} \right) \right]
\end{eqnarray*}
and near the $s=-\frac{1}{2}$ pole we find:
\begin{eqnarray*}
\Delta E_1(\phi_K)&=& \frac{\hbar \lambda}{\sqrt{\pi}m^2}
\lim_{\varepsilon \rightarrow 0}\left( \frac{ \mu^2}{m^2}
\right)^{\varepsilon}
\frac{\Gamma(\varepsilon)}{\Gamma(-\frac{1}{2}+\varepsilon)}
\left[ \frac{2}{3^\varepsilon} \, {}_2
F_1[{\textstyle\frac{1}{2}},\varepsilon,
{\textstyle\frac{3}{2}},-{\textstyle\frac{1}{3}}]-
\frac{1}{(-\frac{1}{2}+\varepsilon)\,4^{-\frac{1}{2}+\varepsilon}}
\right]\\ &=& \frac{\hbar\lambda}{2\pi m^2} \lim_{\varepsilon
\rightarrow 0} \left[ -\frac{3}{\varepsilon}-3
\ln \frac{\mu^2}{m^2}+2 +\ln \frac{3}{4}
-{}_2F_1'[{\textstyle\frac{1}{2}},0,{\textstyle\frac{3}{2}},-{\textstyle\frac{1}{3}}]
+{\cal O}(\varepsilon) \right] \quad ,
\end{eqnarray*}
where ${}_2 F_1'$ is the derivative of the Gauss hypergeometric
function with respect to the second argument. The regularized mass renormalization contribution
to the kink energy is:
\begin{eqnarray*}
\Delta E_2(\phi^K) &=& -\frac{\hbar\lambda}{m^2}
\lim_{s\rightarrow -\frac{1}{2}} \left( \frac{\mu}{m}
\right)^{2s+1}\frac{\Gamma (s+1)}{\Gamma (s)} \lim_{l\to\infty}\frac{6}{l}\zeta_{K_0}(s+1) + {\cal O}(\hbar^2 \frac{\lambda^2}{m^4})\\ &=&-\frac{3\hbar \lambda}{\sqrt{2\pi}m^2}\lim_{s\to
-\frac{1}{2}}\left(\frac{\mu}{m}\right)^{2s+1}\frac{1}{4^{s+\frac{1}{2}}}\frac{\Gamma
(s+\frac{1}{2})}{\Gamma (s)}+{\cal O}(\hbar^2 \frac{\lambda^2}{m^4}) \quad .
\end{eqnarray*}
Near the $s=-\frac{1}{2}$ pole we have:
\begin{eqnarray*}
\Delta E_2 (\phi_K)&=&-\frac{3 \hbar\lambda}{\sqrt{2\pi}m^2}
\lim_{\varepsilon \rightarrow 0} \left( \frac{2 \mu^2}{m^2}
\right)^\varepsilon \frac{4^{-\varepsilon}
\Gamma(\varepsilon)}{\Gamma(-\frac{1}{2}+\varepsilon)}+{\cal O}(\hbar^2\frac{\lambda^2}{m^4})\\ &=&\frac{3\hbar
\lambda}{2\pi m^2} \lim_{\varepsilon \rightarrow 0} \left[
\frac{1}{\varepsilon}+ \ln \frac{\mu^2}{m^2}-\ln 4+\psi(1)-\psi(-\frac{1}{2})+{\cal O}(\varepsilon)
\right]+{\cal O}(\hbar^2 \frac{\lambda^2}{m^4})
\end{eqnarray*}
Again ,the sum of these two contributions leaves a finite remainder and we obtain:
\begin{equation}
\Delta\tilde{E}(\psi_K)=\Delta\tilde{E}_1(\psi_K)+\Delta\tilde{E}_2(\phi_K)=\frac{\hbar m}{2\sqrt{6}}-\frac{3\hbar m}{\pi\sqrt{2}} \label{lpolms} \quad ,
\end{equation}
the same answer as that discovered by regularizing the ultraviolet divergences with a cutoff in the number of modes. The one-loop renormalized kink mass is:
\[
\tilde{E}(\psi_{\rm 1-loop})=\tilde{E}(\psi_{\rm K})+\Delta\tilde{E}(\psi_{\rm K})=\frac{4}{3}
\frac{m^3}{\sqrt{2}\lambda}\left[1-\hbar\frac{\lambda}{m^2}\left(\frac{9}{4\pi}-\frac{\sqrt{3}}{8}\right)+
{\cal O}(\hbar^2 {\textstyle\frac{\lambda^2}{m^4}})\right]=\frac{4}{3}\frac{M^3}{\sqrt{2}\lambda} \quad .
\]
Like in the sine-Gordon model we find a similar pathological situation: when $\hbar\frac{\lambda}{m^2}>\frac{1}{\frac{9}{4\pi}-\frac{\sqrt{3}}{8}}$ $\tilde{E}(\psi_{\rm 1-loop})<0$ and
$M^3<0$. The real cubic root of this negative value of $M^3$ is negative such that
$M$ is also negative suggesting that this model is quantum mechanically unsound in this regime.
\end{itemize}

\section{The heat kernel expansion and the one-loop kink mass shift}

Both when we regularize the DHN formula by means of a cutoff in the number of modes and when we use the zeta
function regularization procedure we must have  a complete knowledge of the spectrum of the $K$ Hessian operator governing
the small kink fluctuations. For generic kinks this information is not available, and we need an alternative procedure to compute the one-loop mass shift
other than the sG and $\lambda\phi^4$ kink mass shifts.
\subsection{The asymptotic expansion of the kink heat function}

To perform this task, we shall rely on the asymptotic expansion of the heat-kernel expansion, which encodes all the spectral information about the $K$ differential operator. The kink spectral function ${\rm Tr}_{L^2}e^{-\beta K}$ admits an integral kernel representation. If $f_k(x)$ are the eigenfunctions of $K$ with eigenvalues $\omega^2(k)$ labeled by a continuous index $k$, the integral
\begin{equation}
K_{K}(x,y;\beta)= \int \! dk \, f_k^{*} (y) \, f_k(x) \, e^{-\beta \omega^2(k)}
\label{integralkernel}
\end{equation}
is the $K$-heat equation kernel. The $K$-heat function is thus:
\begin{equation}
h_{K}(\beta) \equiv {\rm Tr}_{L^2}\, e^{-\beta K}=
\int_{-\infty}^\infty \hspace{-0.1cm} dx \, K_{K}(x,x;\beta) \quad .
\label{heatfunction2}
\end{equation}
The $K_0$-heat equation kernel is found immediately:
\[
K_{K_0}(x,y;\beta)=\frac{1}{\sqrt{4\pi \beta\, }} \,
e^{-\beta v^2}\, e^{- \frac{(x-y)^2}{4 \, \beta\,}}
\]
because in this case: $f_k(x)=\frac{1}{\sqrt{2 \pi}} e^{ikx}$ and $\omega^2(k)=k^2+v^2, k\in{\mathbb R}$.

To estimate $K_{K}(x,y;\beta)$ in the cases where the eigenfunctions of $K$ are not known we shall use the fact that
this integral kernel is the fundamental solution of the $K$-heat equation: the solution of
\begin{equation}
\left[ \frac{\partial}{\partial \beta}-\frac{\partial^2}{\partial
x^2}+v^2 + V(x) \right] K_{K}(x,y;\beta)=0 \label{heateq}
\end{equation}
with a $\delta$-source on the diagonal of ${\mathbb R}^2$ at infinite temperature:
\begin{equation}
K_{K}(x,y;0)=\delta(x-y) \label{heatic} \quad .
\end{equation}
We shall assume the convolution ansatz
\begin{equation}
K_{K}(x,y;\beta)=K_{K_0}(x,y;\beta) \, C(x,y;\beta)
\label{facto}
\end{equation}
in order to benefit from the knowledge of $K_{K_0}(x,y;\beta)$. Plugging  (\ref{facto}) into the $K$-heat equation (\ref{heateq}) we find the transfer equation for $C(x,y;\beta)$:
\begin{equation}
\left( \frac{\partial}{\partial \beta}+\frac{x-y}{\beta}
\frac{\partial}{\partial x}-\frac{\partial^2}{\partial
x^2}+V(x) \right) C(x,y;\beta)=0 \label{heateq2} \quad .
\end{equation}
Moreover, the infinite temperature condition (\ref{heatic}) prompts the condition
\begin{equation}
C(x,y;0)=1 \label{heatic2} \quad .
\end{equation}
The idea is now to solve (\ref{heateq2}) by means of a series expansion.  Firstly, $C(x,y;\beta)$ is expanded around the point $\beta=0$ as a power series in $\beta$:
\begin{equation}
C(x,y;\beta) = \sum_{n=0}^\infty c_n(x,y) \, \beta^n  \quad .
\label{expansion}
\end{equation}
The zero-order density is fixed by the the infinite temperature condition (\ref{heatic2}):
\begin{equation}
c_0(x,y)=1 \quad .
\label{heatic3}
\end{equation}
 Secondly, (\ref{expansion}) is plugged into (\ref{heateq2}), which trades this PDE for the recurrence relations between the densities $c_n(x,y)$ and their first- and second-order derivatives:
\begin{equation}
(n+1) \, c_{n+1}(x,y)+(x-y) \frac{\partial c_{n+1}(x,y)}{\partial
x}+V(x) c_n(x,y)=\frac{\partial^2 c_n(x,y)}{\partial
x^2} \label{recursive1} \quad .
\end{equation}
The solution of (\ref{recursive1}) provides a series expansion of the integral kernel in the form: $K_{K}(x,y;\beta)=K_{K_0}(x,y;\beta) \,\sum_{n=0}^\infty c_n(x,y) \, \beta^n$. In turn, the $K$-heat function
is also expanded in powers of $\beta$
\begin{equation}
h_{K}(\beta) = \frac{e^{-\beta v^2
}}{\sqrt{4 \pi \beta\,}} \sum_{n=0}^\infty {c}_n(K) \,
\beta^n
\label{heatasymptotic} \quad ,
\end{equation}
where the $c_n(K)$ are the Seeley coefficients of the $K$-heat function high-temperature expansion:
\begin{equation}
{c}_n(K)=\int_{-\infty}^\infty  dx \, c_n(x,x) \quad .
\label{seeleycoef}
\end{equation}
The formula (\ref{seeleycoef}) above involves the limit of the densities $c_n(x,y)$ as $y$ approaches $x$, i.e.,
\[
c_n(x,x)=\lim_{y\rightarrow x} c_n(x,y) \quad .
\]
Moreover, derivatives of the densities enter the recurrence relations (\ref{recursive1}). It is thus convenient
to define the $y\to x$ limit of the derivatives
\begin{equation}
{^{(k)}C}_n(x)=\lim_{y \rightarrow x} \frac{\partial^k
c_n(x,y)}{\partial x^k}
\label{newcoef}
\end{equation}
such that
\[
c_n(x,x)={^{(0)}C}_n(x) \quad .
\]
The infinite temperature conditions (\ref{heatic3}) determine the densities ${^{(k)}C}_0(x)$:
\[
{^{(k)} C}_0(x)=\lim_{y\rightarrow x} \frac{\partial^k
c_0}{\partial x^k}= \delta^{k0}
\]
Taking the $k$-th derivative of expression (\ref{recursive1}) with respect to $x$ and later taking the limit $y\rightarrow x$, we obtain recurrence relations between all the densities ${^{(k)}C}_n(x)$:
\begin{equation}
{^{(k)} C}_n(x) =\frac{1}{n+k} \left[ \rule{0cm}{0.6cm} \right.
{^{(k+2)} C}_{n-1}(x) - \sum_{j=0}^k {k \choose j}
\frac{\partial^j V}{\partial x^j}\, \, {^{(k-j)}
C}_{n-1}(x) \left. \rule{0cm}{0.6cm} \right]
\label{capitalAcoefficients}
\end{equation}
affording us the identification of ${^{(k)} C}_n(x)$ in a recursive way from ${^{(k)} C}_0(x)= \delta^{k0}$. Note that the computation of the densities ${^{(k)} C}_n(x)$ requires the densities ${^{(0)} C}_{n-1}(x)$, ${^{(1)} C}_{n-1}(x)$, $\dots$, ${^{(k+2)} C}_{n-1}(x)$, which in turn are determined from the densities ${^{(0)} C}_{n-2}(x)$, ${^{(1)} C}_{n-2}(x)$, $\dots$ ${^{(k+4)} C}_{n-2}(x)$ and so on until we reach the densities ${^{(k)} C}_0(x)= \delta^{k0}$. Therefore, the evaluation of the density ${^{(k)} C}_n(x)$ is based on the densities that appear in the following pyramid:
\[
\begin{array}{cccccccc}
{^{(0)} C}_n(x) & & & & & & & \\
{^{(0)} C}_{n-1}(x) & {^{(1)} C}_{n-1}(x) & {^{(2)} C}_{n-1}(x) & & & & & \\
{^{(0)} C}_{n-2}(x) & {^{(1)} C}_{n-2}(x) & {^{(2)} C}_{n-2}(x) & {^{(3)} C}_{n-2}(x) & {^{(4)} C}_{n-2}(x) & & & \\ \dots & \dots & \dots & \dots & \dots & \dots & & \\
{^{(0)} C}_{0}(x) & {^{(1)} C}_{0}(x) & {^{(2)} C}_{0}(x) & {^{(3)} C}_{0}(x) & {^{(4)} C}_{0}(x) & \dots & {^{(2n-1)} C}_{0}(x) & {^{(2n)} C}_{0}(x) \quad ,
\end{array}
\]
i.e., $(n+1)^2$ densities ${^{(j)} C}_m(x)$ must be evaluated previously.

We list the first four densities $c_n(x,x)$ for the kink fluctuation operator in a general (1+1) dimensional one-component scalar field theoretical model:
\begin{eqnarray*}
c_0(x,x)={^{(0)} C}_0(x)&=&1 \\
c_1(x,x)={^{(0)} C}_1(x)&=&-V(x)  \\
c_2(x,x)={^{(0)} C}_2(x)&=&-\frac{1}{6} \, \frac{\partial^2 {V}}{\partial
x^2}+\frac{1}{2} \, (V(x))^2 \\
c_3(x,x)={^{(0)} C}_3(x)& =& -\frac{1}{60} \frac{\partial^4 V}{\partial
x^4} +\frac{1}{6} V(x) \frac{\partial^2
V}{\partial x^2} + \frac{1}{12} \frac{\partial
V}{\partial x} \frac{\partial V}{\partial
x}-\frac{1}{6} {V(x)}^3 \quad .
\end{eqnarray*}
There is an interesting point about these densities: they are the infinite conserved charges of the KdV
equation, see e.g. \cite{IGA}. Consider the family of differential
operators
\[
K(\beta)=-\frac{\partial^2}{\partial x^2}+v^2+V(x,\beta),
\]
where the family of potentials solve the KdV equation:
\[
\frac{\partial V}{\partial\beta }-6V\frac{\partial V}{\partial x}+\frac{\partial^3V}{\partial x^3}=0 \qquad .
\]
The $\beta$-evolution of $K(\beta)$ can be expressed in the Lax pair
form
\[
\frac{\partial K}{\partial\beta}+[K,M]=0 \qquad , \qquad
M(\beta)=4\frac{\partial^3}{\partial
x^3}+3\left(V\frac{\partial }{\partial x}+\frac{\partial }{\partial x}V\right)+B(\beta)
\]
such that it is iso-spectral:
\[
K(\beta)=g(\beta)K(0)g^{-1}(\beta) \, \, , \quad M(\beta)=\frac{\partial g}{\partial\beta}g^{-1}(\beta) \quad .
\]
 The densities codify the
spectrum of $K$. Ergo,
\[
{\rm Tr}_{L^2} e^{-\int_0^\beta \, d\beta^\prime \, K(\beta^\prime)}= {\rm Tr}_{L^2} e^{-\beta  K(0)}=
      \int \, dx \, \frac{e^{-\beta v^2}}{\sqrt{4\pi\beta}}\sum_{n=0}^\infty \, c_n(x,x;\beta)\beta^n \, \Rightarrow \, c_n(x,x;\beta)\neq f(\beta) \quad .
\]

\subsection{The asymptotic high-temperature expansion of the quantum correction}

The $K$-heat function $h_{K^\bot}(\beta)$, basic in the computation of the one-loop mass shift via zeta function regularization, admits the series expansion:
\begin{equation}
h_{K^\bot}(\beta)= \frac{e^{-\beta {v}^2}}{\sqrt{4\pi}} {c}_0(K)\beta^{-\frac{1}{2}} + \frac{e^{-\beta {v}^2}}{\sqrt{4\pi}} {c}_1(K)\beta^{\frac{1}{2}} + \frac{e^{-\beta {v}^2}}{\sqrt{4\pi}} \sum_{n=2}^\infty {c}_n(K)\beta^{n-\frac{1}{2}}-1 \qquad .
\label{heatfunction3}
\end{equation}
The first two Seeley coefficients are:
\[
c_0(K)=\int_{-\frac{l}{2}}^\frac{l}{2} c_0(x,x)dx = l \hspace{0.5cm} \mbox{and} \hspace{0.5cm} c_1(K)= \lim_{l\to\infty}\int_{-\frac{l}{2}}^\frac{l}{2} c_1(x,x)dx = -\int_{-\infty}^\infty V(x) dx = -\left< V(x) \right> \quad .
\]
The zero-point energy renormalization
\begin{equation}
h_{K^\bot}(\beta) - h_{K_0} (\beta) = \frac{e^{-\beta {v}^2}}{\sqrt{4\pi}} {c}_1(K)\beta^{\frac{1}{2}} + \frac{e^{-\beta {v}^2}}{\sqrt{4\pi}} \sum_{n=2}^\infty {c}_n(K)\beta^{n-\frac{1}{2}}-1
\label{heatseries}
\end{equation}
 cancels the (divergent at the $l=\infty$ limit) contribution to the heat function of the $c_0(K)$ coefficient.

Another crucial cancelation, known as the heat kernel renormalization criterion, see e.g. \cite{Bordag2002}, occurs between the contribution of the $c_1(K)$ to the one-loop mass shift and the mass renormalization energy:
\begin{eqnarray*}
\bigtriangleup E_1^{(1)}(\phi_K)[s]&=&\frac{\hbar\gamma_d^2}{2}\left(\frac{\mu^2}{m_d^2}\right)^{s+\frac{1}{2}}
\frac{c_1(K)}{\sqrt{4\pi}}\frac{1}{\Gamma(s)}\int_0^\infty \, d\beta \, \beta^{s-\frac{1}{2}}e^{-\beta v^2}\\
&=& -\frac{\hbar\gamma_d^2}{2}\left(\frac{\mu^2}{m_d^2}\right)^{s+\frac{1}{2}}
\frac{\langle V(x) \rangle}{\sqrt{4\pi}}\frac{\Gamma(s+\frac{1}{2})}{\Gamma(s)}\frac{1}{v^{2s+1}}\\
\bigtriangleup E_2(\phi_K)[s]&=&\frac{\hbar\gamma_d^2}{2}\left(\frac{\mu^2}{m_d^2}\right)^{s+\frac{1}{2}}
\frac{\langle V(x) \rangle}{\sqrt{4\pi}}\frac{\Gamma(s+\frac{1}{2})}{\Gamma(s)}\frac{1}{v^{2s+1}} \quad .
\end{eqnarray*}
Therefore, the divergences in the $s=-\frac{1}{2}$ pole are canceled and we finish with the following one-loop mass shift formula derived from the high-temperature expansion of the $K$- and $K_0$-heat functions:
\begin{equation}
\bigtriangleup E(\phi_K)= \frac{\hbar \gamma_d^2}{2} \lim_{s\rightarrow -\frac{1}{2}} \frac{1}{\Gamma(s)} \int_0^\infty \hspace{-0.3cm} d\beta \,\beta^{s-1} \left[ \frac{e^{-\beta {v}^2}}{\sqrt{4\pi}} \sum_{n=2}^\infty {c}_n(K)\beta^{n-\frac{1}{2}}-1\right] \quad .
\label{computation1}
\end{equation}

\section{A computational treatment of the one-loop kink mass shift formula}

\subsection{The computational formula}

In this Section we shall convert the one-loop kink mass shift formula (\ref{computation1}) into a formula that will afford us a precise estimation of the quantum correction in a general model simply by using a desk computer. In this computational scenario the series (\ref{computation1}) is first truncated at a certain order $N$. Before doing all this we note the
following behaviour at high and low temperatures of the difference of the $K$- and $K_0$-heat functions:
\[
\lim_{\beta \rightarrow 0} \left(h_{K^\bot}(\beta)-h_{K_0}(\beta)\right)=-1 \hspace{1cm}; \hspace{1cm} \lim_{\beta \rightarrow \infty} \left(h_{K^\bot}(\beta)-h_{K_0}(\beta)\right)=0 \quad .
\]
We shall denote as $S_{K}(\beta,N)$ the truncation of the series expansion (\ref{heatseries}) of this quantity:
\begin{equation}
S_{K}(\beta,N) =  \frac{e^{-\beta {v}^2}}{\sqrt{4\pi}} {c}_1(K)\beta^{\frac{1}{2}} + \frac{e^{-\beta {v}^2}}{\sqrt{4\pi}} \sum_{n=2}^N {c}_n(K)\beta^{n-\frac{1}{2}}-1 \quad .
\label{truncated}
\end{equation}
This truncation required for computational effectiveness involves a troublesome effect. The asymptotic behaviours of the series (\ref{heatseries}) and the truncated series (\ref{truncated}) for $h_{K}(\beta)-h_{K_0}(\beta)$ do not coincide at low temperature. It is easy to check that:
\[
\lim_{\beta \rightarrow 0} S_{K}(\beta,N)= -1\hspace{1cm}; \hspace{1cm}\lim_{\beta \rightarrow \infty} S_{K}(\beta,N)= -1 \quad .
\]
 As expected, the behaviour of $h_{K^\bot}(\beta)-h_{K_0}(\beta)$ and $S_{K}(\beta,N)$ fits extremely well in the high-temperature regime, small values of $\beta$. If $N$ is large, the identification of the series and the partial sum can be continued up to a value $\beta=\beta_0(N)$ such that $h_{K^\bot}(\beta_0(N))-h_{K_0}(\beta_0(N))$ is close to zero. The convergence of the partial sums to the series implies that $\lim_{N\rightarrow \infty} \beta_0(N) =\infty$. This is depicted in Figure 2 for the $\lambda\phi^4$ model and several values of $N$.
 \begin{figure}[ht]
\centerline{\includegraphics[height=3cm]{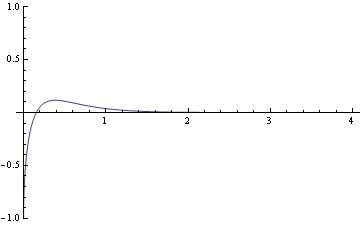}\hspace{2cm}
\includegraphics[height=3cm]{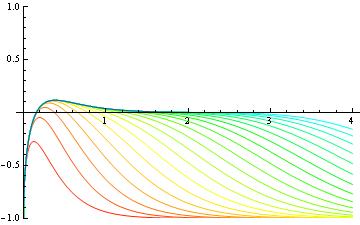} }
\caption{\textit{$\lambda\phi^4$ kink:  function  $h_{K^\bot}(\beta)-h_{K_0}(\beta)$ (left) partial sums $S_{K}(\beta,N)$ for growing values of $N$ from the bottom (red) to the top (violet) line (right), both depicted as functions of $\beta$.}}
\end{figure}
Replacing $h_{K^\bot}(\beta)-h_{K_0}(\beta)$ by the partial sum $S_{K}(\beta,N)$ in formula (\ref{computation1})
we obtain:
\[
\Delta E(\phi_K)=\Delta E(\phi_K;N)+R(\beta_0,N)= \frac{\hbar\gamma_d^2}{2} \lim_{s\rightarrow -\frac{1}{2}} \frac{1}{\Gamma(s)} \int_0^{\beta_0} \hspace{-0.3cm} d\beta \,\left[ \frac{e^{-\beta {v}^2}}{\sqrt{4\pi}} \sum_{n=2}^N {c}_n(K)\beta^{s+n-\frac{3}{2}}-\beta^{s-1} \right] + R(\beta_0,N)
\]
where $R(\beta_0,N)$ is a negligible contribution such that:
\[
\lim_{N\rightarrow \infty} R(\beta_0(N),N) =0 \quad .
\]
The next steps are to perform the $\beta$ integration and to take the $s\to -\frac{1}{2}$ limit:
\begin{eqnarray*}
\Delta E(\phi_K;N) &=&
\frac{\hbar\gamma_d^2}{2} \lim_{s\rightarrow -\frac{1}{2}} \frac{1}{\Gamma(s)} \left[ \frac{1}{\sqrt{4\pi}}\sum_{n=2}^N {c}_n(K)(v^2)^{\frac{1}{2}-n-s}\gamma[-\frac{1}{2}+n+s,\beta_0 v^2]- \frac{\beta_0^s}{s}\right]  =\\ &=&
-\frac{\hbar\gamma_d^2}{2\sqrt{4\pi}} \left[ \frac{1}{\sqrt{4\pi}}\sum_{n=2}^N {c}_n(K)(v^2)^{1-n}\gamma[n-1,\beta_0 v^2]+\frac{2}{\sqrt{\beta_0}}\right]\quad ,
\end{eqnarray*}
where $\gamma[-\frac{1}{2}+n+s,\beta_0 v^2]$ are incomplete Euler Gamma functions. Finally,
\begin{equation}
\Delta E(\phi_K;N)= \hbar \gamma_d^2 \left[ -\frac{1}{8\pi} \sum_{n=2}^N {c}_n(K)(v^2)^{1-n}\gamma[n-1,\beta_0 v^2]-\frac{1}{2\sqrt{\pi\beta_0}} \right]
\label{numericalcorrection}
\end{equation}
is the basic formula for our computational approach.

\subsection{Constructing a symbolic algorithm}

We are now ready to construct a program by means of a symbolic software application such as Mathematica. The key process is to implement formula (\ref{numericalcorrection}) in this algorithm. In order to obtain an estimation of $\Delta E(\phi_K;N)$ we must compute the $c_n(K)$, $n=1,2,\dots, N$, Seeley coefficients. This computation requires the integration of the $c_n(x,x)$, $n=1,2,\dots, N$, densities over the whole real line. Previously, we need to identify $(N+1)^2$ densities ${}^{(k)}C_n(x)$ using the recurrence relations (\ref{capitalAcoefficients}) in order to evaluate $c_N(x,x)={}^{(0)}C_N(x)$. For instance, the calculation of $\Delta E(\phi_K;20)$ demands the computation of 441 densities of this type. Once the Seeley coefficients have been identified we apply formula (\ref{numericalcorrection}) in a direct way. All these tasks are divided into three subroutines described below:

\begin{itemize}
\item Calculation of the $c_n(x,x)$ densities.

 A key point in our symbolic program is the possibility of expressing all the densities as functions of the $\phi_K$ field without the need to use their explicit dependence on $x$. Once we identify the ${}^{(k)}C_{n-1}(x)$ as functions of $\phi_K(x)$, the next-order densities ${}^{(k)}C_{n}(x)$ can also be written as functions of the field. The only elements in the recurrence relations (\ref{capitalAcoefficients}) that we need to determine in terms of $\phi_K$ are the derivatives $\frac{d^j V}{d x^j}$ . To do this we recall that $V(\phi_K)=\frac{\partial^2 U}{\partial \phi^2}[\phi_K]-v^2$ and we use the first order ODE (\ref{ode1}):
 \begin{eqnarray*}
\frac{dV}{dx}&=&\frac{dV}{d\phi}[\phi_K]\frac{d\phi_K}{dx}= \frac{\partial^3 U}{\partial \phi^3}[\phi_K] \sqrt{2U[\phi_K]} \\ \frac{d^2 V}{d x^2}&=& \left(\frac{\partial^4 U}{\partial\phi^4}[\phi_K]\sqrt{2U[\phi_K]}+\frac{\partial U^3}{\partial\phi^3}[\phi_K]\frac{1}{\sqrt{2U[\phi_K]}}\frac{\partial U}{\partial\phi}[\phi_K]\right)\frac{d\phi_K}{d x}\\ &=& 2\frac{\partial^4 U}{\partial\phi^4}[\phi_K]U[\phi_K]+
\frac{\partial U^3}{\partial\phi^3}[\phi_K]\frac{\partial U}{\partial\phi}[\phi_K]
\end{eqnarray*}
Higher than second-order derivatives $\frac{d^j V}{d x^j}, j\geq 3$, can be expressed as functions of $\phi_K$ by following the same pattern. A symbolic program easily manages this computational strategy in each step from $n-1$ to $n$ in order to calculate the ${}^{(k)}C_{n}(\phi_K)$ densities in an optimum way. For instance, we find that:
\begin{eqnarray*}
c_1(x,x)&=& v^2 - \frac{\partial^2 U}{\partial \phi^2}[\phi_K(x)]\\
c_2(x,x)&=& \frac{1}{2} \left( v^2 - \frac{\partial^2 U}{\partial \phi^2}[\phi_K(x)] \right)^2 - \frac{1}{6} \frac{\partial U}{\partial \phi}[\phi_K(x)] \frac{\partial^3 U}{\partial \phi^3}[\phi_K(x)] - \frac{1}{3} U[\phi_K(x)] \frac{\partial^4 U}{\partial \phi^4}[\phi_K(x)
]
\end{eqnarray*}
and so on. The following Mathematica code

\vspace{0.2cm}

\hspace{1cm}\begin{minipage}{14cm}
\texttt{\footnotesize
densitycoefficients[potential\_, vacuum1\_, vacuum2\_, nmax\_] :=
  Module[\{var4, var5, var6, tomax, d1, v, v0, oper, f6, x7, coeficientes,
    alfa, coa, k8, co\}, (var4[ph1\_] = potential/.\{y -> ph1\}; var6[ph1\_] = Simplify[PowerExpand[Sqrt[2 var4[ph1]]]];
    var5[ph1\_] = Sign[var6[1/2(vacuum1+vacuum2)]] var6[ph1];
    coeficientes = \{\};
    v[x\_] = Simplify[(D[var4[ph1],\{ph1,2\}]) /.\{ph1 -> ph1[x]\}];
    v0[x\_] = Simplify[(D[var4[ph1],\{ph1,2\}])/.\{ph1 -> vacuum1\}];
    d1[fun\_] := Simplify[(D[fun,x]) /. \{ ph1'[x]-> var5[ph1[x]]\}];
    oper[fu8\_, n1\_] :=
     Simplify[Nest[f6, x7, n1] /. \{f6 -> d1, x7 -> fu8\}];
    tomax = 2 nmax;
    For[alfa = 0, alfa < tomax + 0.5, coa[0, alfa] = 0; alfa++];
    coa[0, 0] = 1;
    co[k5\_, alfa\_] :=
    Simplify[1/(k5 + alfa) (coa[k5 - 1, alfa + 2] - $\sum_{\rm r5 = 0}^{{\rm alpha}}$ Binomial[
            alfa, r5] oper[v[x] - v0[x], r5]  coa[k5 - 1,
            alfa - r5] ) ];
    For[k8 = 1, k8 < nmax + 0.5, tomax = tomax - 2;
     For[alfa = 0, alfa < tomax + 0.5, coa[k8, alfa] = co[k8, alfa];
      		If[alfa == 0, coeficientes = Append[coeficientes, coa[k8, 0]]];
      		alfa++];
     	k8++]; Return[coeficientes])];
}\end{minipage}

\vspace{0.2cm}

\noindent defines the module \texttt{densitycoefficients[potential\_, vacuum1\_, vacuum2\_, nmax\_]} that is capable of performing this work. The arguments of this computational function are \texttt{potential}, the $U(y)$ potential written by prescription as a function of the $y$ variable, \texttt{vacuum1} and $\texttt{vacuum2}$, the two vacua connected by the kink solution in increasing order, and \texttt{nmax}, the $N$-order truncation chosen in the computation of $\Delta E(\phi_K; N)$ .

\item Calculation of the Seeley coefficients.

We proceed to the calculation of the $c_n(K)$ Seeley coefficients by integrating the $c_n(x,x)$ densities over the whole real line. A novel strategy to perform such integrations will reduce the main difficulty; namely, to cope symbolically or numerically with infinite integration domains. Recall that the densities $c_n(x,x)$ are functions of the kink field $\phi_K$: $c_n(x,x)=f_n(\phi_K)$. The change of variables $z=\phi_K(x)$ is a bijective function from $[-\infty,\infty]$ to $[z_1=\phi^{(1)}, z_2=\phi^{(2)}]$ such that:
\[
c_n(K)=\int_{-\infty}^\infty f_n(\phi_K(x)) dx =\int_{z_1}^{z_2} \frac{f_n(z)}{\sqrt{2U(z)}} dz
\]
 because $dz=\frac{d\Phi_K}{dx}dx =\sqrt{2U(\phi_K)}dx$. This manoeuvre avoids the problem of evaluating an integration over an infinite domain and explicitly shows the topological character of the Seeley coefficients, which only depend on the values of certain functions of the field -let us call it generalized superpotentials- at the  vacuum points. The following Mathematica module called \texttt{seeleycoefficients[potential\_, vacuum1\_, vacuum2\_, nmax\_]}

\vspace{0.2cm}

\hspace{1cm}\begin{minipage}{14cm}
\texttt{\footnotesize
seeleycoefficients[potential\_, vacuum1\_, vacuum2\_, nmax\_] :=
  Module[\{coef, f, f1, f2, a = \{\},
    k8\}, (coef = densitycoefficients[potential, vacuum1, vacuum2, nmax]; f1[y\_] = Simplify[PowerExpand[ Sqrt[2 potential] ]] ; f2[y\_] = Sign[f1[1/2 (vacuum1+vacuum2)]] f1[y];
    For[k8 = 1, k8 < nmax + 0.5, f[y\_]= Simplify[(coef[[x8]]/.\{ph[x]->y\} )/f2[y]];
     a = Append[a,
       Integrate[f[y], \{y, vacuum1, vacuum2\}]]; k8++]; Return[a])];
}\end{minipage}

\vspace{0.2cm}

\noindent  does the trick. Here again the arguments of this function are the potential $U(y)$, the vacuum points $\phi^{(1)}$ and $\phi^{(2)}$ and the truncation order $N$. Note one remarkable fact: there is no need to know the explicit expression of the kink solution !! The potential $U(\phi)$ and the vacua connected by the kink encode all the necessary information in this computation.

From a computational point of view this subroutine is vital for the precision of the final result. Some dysfunctional behavior may arise from this algorithm when the value of the truncation order $N$ is large. This may happen when the densities $c_n(x,x)$ include such large/small factors than some loss of precision occurs. This effect can be detected by checking the sequence of Seeley coefficients. Theoretically these coefficients tend to vanish when the index $n$ increases. Thus, if there is some anomaly in the Seeley coefficient sequence starting in a given $n_0$ the quantum correction estimated as an asymptotic series can be trusted only up to a truncation order $N=n_0$.

\item Estimation of the quantum correction.

It remains to optimize the value of the $\beta_0(N)$ choice, related to the asymptotic behaviour of the partial sums $S_{K}(\beta,N)$. Firstly, we find a point $\beta$ at which $S_{K}(\beta,N)$ becomes approximately $-1$ and, second, we move back until we obtain a value $\beta_0$ such that $S_{K}(\beta,N)$ almost vanishes. With this choice of optimum $\beta_0$ we obtain the quantum correction by applying (\ref{numericalcorrection}), and the task is accomplished by the subroutine \texttt{quantumcorrection[potential\_, vacuum1\_, vacuum2\_, nmax\_]}

\vspace{0.2cm}

\hspace{1cm}\begin{minipage}{15cm}
\texttt{\footnotesize
quantumcorrection[potential\_, vacuum1\_, vacuum2\_, nmax\_] :=
Module[\{var4, vhess1, heat, a, b0,
    corr\}, (var4[ph1\_] = potential /. \{y -> ph1\}; vhess1 = Simplify[(D[var4[ph1],\{ph1,2\}])/.\{ph1 -> vacuum1\}];
    a = seeleycoefficients[potential, vacuum1, vacuum2, nmax];
   heat[b\_] = E\^ {}\{-b vhess1\}/(2 Sqrt[Pi]) ($\sum_{\rm n = 1}^{\rm nmax}$ (a[[n]]
b${}^{\rm n -\frac{1}{2}}$) - 1; b0 = 10;
    While[heat[b0] > -0.9, b0 = b0 + 1];
    For[h = 0.1, h > 0.001, h = h/2, b0 = b0 + 2 h;
     While[Abs[heat[b0]] >= Abs[heat[b0 - h]], b0 = b0 - h]];
    corr = -1/(2 Sqrt[Pi b0]) - 1/(8 Pi) ($\sum_{\rm n=2}^{\rm nmax}$ (a[[n]]
vhess1${}^{-n+ 1}$ (Gamma[n - 1] - Gamma[n - 1, b0 vhess1]) )); Return[corr])];
}\end{minipage}
\end{itemize}
The above Mathematica code characterizes the application that automates the computation of the quantum correction to the kink mass in one-component scalar field theory.

The KinkMassQuantumCorrection\_Lite.nb file containing this Mathematica code can be download at the web page http://campus.usal.es/$\sim$mpg/General/Mathematicatools. We recommend this option in order to avoid transcription errors in the code.

\subsection{Testing the algorithm in the sine-Gordon and $\lambda\phi^4$ models}

\subsubsection{The algorithm and the sine-Gordon kink}

The exact one-loop sG-kink mass shift, calculated by means of the DHN formula regularized either by a cutoff or by the zeta function procedure, is: $\Delta \tilde{E}_{\rm sG}^{(2\pi)}(\phi_K) =-\hbar m (\frac{1}{\pi}) \approx -0.31831 \hbar m$. The result for this magnitude provided by the asymptotic approach is accessible in the symbolic algorithm. Recall that $\gamma_d^2=m$ and that the dimensionless potential is $U(\phi)=1-\cos \phi$. By running the KinkMassQuantumCorrection\_Lite.nb file and then executing the command \texttt{quantumcorrection[$1-\cos ({\rm y})$,0,2$\pi$,nmax]} with $\texttt{nmax}=10,20,30$ the following results
are obtained:
\begin{center}
\begin{tabular}{l}
\texttt{quantumcorrection[$1-\cos ({\rm y})$,0,2$\pi$,10]=-0.319113} \\
\texttt{quantumcorrection[$1-\cos ({\rm y})$,0,2$\pi$,20]=-0.318321} \\
\texttt{quantumcorrection[$1-\cos ({\rm y})$,0,2$\pi$,30]=-0.318310}\quad .
\end{tabular}
\end{center}
The relative errors are: 1) $0.252263\%$, truncation at 10 terms, 2) $0.00345755\%$, truncation at 20 terms, and 3) $0.0000600568\%$, truncation at 30 terms. The final result fits in extraordinarily well with the exact one-loop sG-kink mass shift and it is closer for higher truncation order.

The symbolic algorithm is prepared such that it will provide not only the final outcome but also offer the intermediate quantities needed in the computation.  For instance, the sG-kink Seeley coefficients are accessible by executing the command \texttt{seeleycoefficients[$1-\cos ({\rm y})$,-1,1,30]}. See the list of the 30 first coefficients $\{c_n(K)\}_{n=1}^{30}$:
\[
\begin{array}{l}
\{4., 2.66667, 1.06667, 0.304762, 0.0677249, 0.0123136, 0.0018944, \
0.000252587, 0.0000297161, \\ \hspace{0.2cm}
3.12801\cdot 10^{-6}, 2.97906\cdot 10^{-7},
2.59049\cdot 10^{-8}, 2.07239\cdot 10^{-9}, 1.5351\cdot 10^{-10}, 1.05869\cdot 10^{-11},\\ \hspace{0.2cm}
6.83027\cdot 10^{-13}, 4.13956\cdot 10^{-14}, 2.36546\cdot 10^{-15}, 1.27863\cdot 10^{-16},
6.55706\cdot 10^{-18}, 3.19857\cdot 10^{-19}, \\ \hspace{0.2cm} 1.48771\cdot 10^{-20}, 6.61203\cdot 10^{-22},
2.81363\cdot 10^{-23}, 1.14842\cdot 10^{-24}, 4.50361\cdot 10^{-26}, 1.69947\cdot 10^{-27},\\ \hspace{0.2cm}
6.17991\cdot 10^{-29}, 2.16839\cdot 10^{-30}, 7.35047\cdot 10^{-32}\}
\end{array}
\quad .
\]
The command \texttt{densitycoefficients[$1-\cos ({\rm y})$,-1,1,N]} unveils the densities $c_n(x,x)$ needed in the definition of the Seeley coefficients (\ref{seeleycoef}). The general form of these densities is:
\[
c_n(x,x)=\frac{2^n}{(2n-1)!!} \sin^2 \frac{\phi_K}{2} = \frac{2^{n-1}}{(2n-1)!!} U(\phi_K)
\]
and, henceforth,
\[
c_n(K)=\frac{2^n}{(2n-1)!!}\int_0^{2\pi}\sin \frac{z}{2}\frac{dz}{2}=\frac{2^{1+n}}{(2n-1)!!} \quad .
\]

\subsubsection{The algorithm and the $\lambda\phi^4$ kink}

The exact DHN one-loop $\lambda\phi^4$-kink mass shift is:
\[
\Delta \tilde{E}_{\phi^4}(\phi_K) =\frac{\hbar m}{\sqrt{2}}\left(\frac{1}{2\sqrt{3}}-\frac{3}{\pi}\right) \approx -0.666255 \frac{\hbar m}{\sqrt{2}}
\]
 Now $\gamma_d^2=\frac{m}{\sqrt{2}}$ and the dimensionless potential is $U(\phi)=\frac{1}{2}(\phi^2-1)^2$. Therefore the expected dimensionless quantum correction to the kink is $-0.666255$. Again, running the KinkMassQuantumCorrection\_Lite.nb file on a Mathematica platform and then executing later the command \texttt{quantumcorrection[$\frac{1}{2} ({\rm y}^2-1)^2$,-1,1,10]} we obtain the estimation of the one-loop quantum correction to the kink mass in the asymptotic approach truncated to 10 terms. The value of $N$ must be chosen adequately so that $N$ will be large enough to obtain a precise answer but not so large that the process becomes very much time consuming. The processing time is very sensitive to the complexity of the potential $U(y)$; note that the symbolic software algorithm must manipulate and simplify the function $U(y)$ and its successive derivatives. Trying three increasing values of $N$, we obtain:
\begin{center}
\begin{tabular}{l}
\texttt{quantumcorrection[$\frac{1}{2} ({\rm y}^2-1)^2$,-1,1,10]=-0.665894} \\
\texttt{quantumcorrection[$\frac{1}{2} ({\rm y}^2-1)^2$,-1,1,20]=-0.666241} \\
\texttt{quantumcorrection[$\frac{1}{2} ({\rm y}^2-1)^2$,-1,1,30]=-0.666254}
\end{tabular}
\quad .
\end{center}
The relative errors are: 1) $0.054112\%$, truncation at 10 terms, 2) $0.00202985\%$ truncation at 20 terms, and 3) $0.0000786421\%$, truncation at 30 terms. These results guarantee the precision of the method. The command \texttt{seeleycoefficients[$\frac{1}{2} ({\rm y}^2-1)^2$,-1,1,30]} prompts the list of the 30 first Seeley coefficients $c_n(K)$, $n=1,\dots,30$, for the $\lambda\phi^4$-kink.
\[
\begin{array}{l}
\{12., 24., 35.2, 39.3143, 34.7429, 25.2306, 15.5208, 8.27702,
3.89498, 1.63998, 0.624754, 0.217306, \\ \hspace{0.2cm} 0.0695378, 0.0206038,
0.00568381, 0.00146679, 0.000355585, 0.0000812766, 0.0000175733, \\\hspace{0.2cm}
3.60478\cdot 10^{-6}, 7.03372\cdot 10^{-7}, 1.3086\cdot 10^{-7}, 2.3264\cdot 10^{-8},
3.95983\cdot 10^{-9}, 6.46503\cdot 10^{-10}, \\ \hspace{0.2cm} 1.01412\cdot 10^{-10}, 1.53075\cdot 10^{-11},
2.22655\cdot 10^{-12}, 3.12498\cdot 10^{-13}, 4.23726\cdot 10^{-14} \}
\end{array}
\quad .
\]
The command \texttt{densitycoefficients[$\frac{1}{2} ({\rm y}^2-1)^2$,-1,1,N]} in turn provides us with the densities $c_n(x,x)$:
\[
c_n(x,x)=\frac{3\cdot 2^n}{(2n-1)!!} (1-\phi_K^2)\left[4^{n-1}-(4^{n-1}-1)\phi_K^2 \right] = \frac{3\cdot 2^{n+1}}{(2n-1)!!} \left[ 4^{n-1} U(\phi_K)-\frac{1}{4}\phi_K U'(\phi_K) \right]
\]
leading to the general form of the $\lambda\phi^4$-kink Seeley coefficients:
\[
c_n(K)=\frac{3\cdot 2^n}{(2n-1)!!}\int^1_{-1}\, dz \, \left(4^{n-1}-(4^{n-1}-1)z^2\right)=\frac{2^{1+n}+8^n}{(2n-1)!!} \quad .
\]

\begin{figure}[ht]
\centerline{\includegraphics[height=3cm]{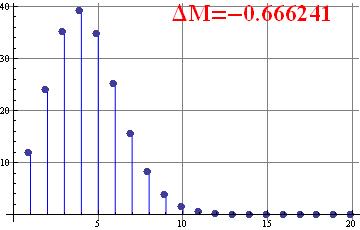}\hspace{1cm}
\includegraphics[height=3cm]{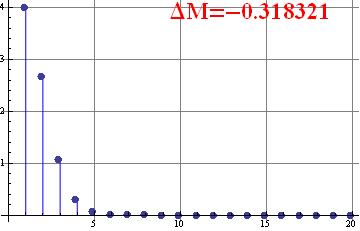} }
\caption{\textit{\small Seeley coefficients depicted as bar charts for the $\phi^4$ model (left) and the sine-Gordon model (right). The one-loop shifts are also shown in red }}
\end{figure}

We now emphasize the strong points in these new calculations.
\begin{itemize}

\item In comparison with the primitive symbolic algorithm used in Reference \cite{Alonso2002}, the present Mathematica program has been extraordinarily refined. The error committed with respect to the exact result in the one-loop sG-kink mass shift truncated at $N=10$ was 6.00 percent and the relative error committed in the present calculation is 0.25 percent. In the $\lambda\phi^4$ kink case, the past and present errors are 0.07 and 0.05 percent. It is the better fitting of the parameter $\beta_0$ in the sG-kink that helps to greatly improve the approximation. Also, integrating in the field space rather than in $x$ is more beneficial in the sG kink
    because the exponential tails of the kink solution decay more slowly to the vacuum values than their counterparts in the $\lambda\phi^4$ kink.

 \item Skipping the integration over infinite domains also helps to compute many more coefficients. A frequent criticism of the Gilkey-de Witt procedure, the difficulty of computing many coefficients, is thus circumvented in this relatively easy problem. Needless to say, the errors are hugely reduced.

  \item In the basic sG and $\lambda\phi^4$ kinks, the functional dependence on the kink profile of the Seeley densities is independent of $n$. This is not so, as we shall see, in other cases, hindering the calculation of the Seeley coefficients.

\end{itemize}

\section{Other models: one-loop corrections to the mass of exotic kinks}

It is possible, in principle, to apply the algorithm to compute the one-loop kink mass shifts in any other one-component scalar field theory model. The inputs that must be introduced in the algorithm are very few: the (dimensionless) potential supporting kinks; the nearest neighbors vacuum points connected by the kink that must belong to a single point in the vacuum moduli space, and the truncation order. We stress that the choice of an excessively large value of $N$ can finally cause a loss of precision, a feature characteristic of asymptotic series.

\subsection{One-loop kink mass corrections in distinguished models}

In Table 1 we assemble the inputs necessary to determine the kink masses up to one-loop order as well as the one-loop kink
mass shifts and truncation order $N$ in several models discussed in the Literature.

\begin{table}[ht]
\hspace{1cm}\begin{tabular}{|c||c|c|c|c|c|}
\hline
\rule[-0.3cm]{0.cm}{0.8cm} $U(\phi)$ & Vacuum $\phi^{(1)}$ & Vacuum $\phi^{(2)}$ & Classical Energy & $\Delta \tilde{E}/\hbar m_d $ & $N$ \\ \hline
\rule[-0.3cm]{0.cm}{0.8cm} $1-\cos (2\phi)$ & $0$ & $\pi$ & 4& $-0.636894$ & 20 \\ \hline
\rule[-0.3cm]{0.cm}{0.8cm} $\frac{1}{4}(\sinh^2\phi-1)^2$ & $-{\rm arcsinh}\, 1 $ & ${\rm arcsinh}\, 1 $ & $\frac{3}{\sqrt{2}}{\rm arcsinh}\, 1-1$ & $-0.73433$ & 17 \\ \hline
\rule[-0.3cm]{0.cm}{0.8cm} $\frac{1}{2}(\phi^2+1)(\phi^2-1)^2$ & $-1$ & $1$ & $\frac{\sqrt{2}}{4} + \frac{5}{4} {\rm arcsinh}\, 1$ & $-1.10078$ & 16 \\ \hline
\rule[-0.3cm]{0.cm}{0.8cm}  $\frac{1}{2}(\phi^4-1)^2$ & $-1$ & $1$ & $\frac{8}{5}$ & $-1.89063$ & 16 \\ \hline
\rule[-0.3cm]{0.cm}{0.8cm} $\frac{1}{2}(\phi^8-1)^2$ & $-1$ & $1$ & $\frac{8}{9}$  & $-6.22270$ & 16 \\ \hline
\rule[-0.3cm]{0.cm}{0.8cm} $\frac{8}{9}(\frac{1}{4}-\phi^2)^2(1-\phi^2)^2$ & $-\frac{1}{2}$ & $\frac{1}{2}$ & $\frac{19}{90}$ & $-0.320753$ & 9 \\ \hline
\end{tabular}
\caption{The one-loop kink mass shift in several one-component scalar field theory models.}
\end{table}
The first model is simply the sine-Gordon model with a re-scaling of the scalar field: $\phi\to 2\phi$.  In this re-scaled sine-Gordon model the distance between two nearest neighbors vacuum points is only $\pi$. Therefore, we have a very good test to the algorithm because a simple change of variables unveils the exact relationship between the re-scaled sG kink and the proper sG kink mass shifts. One easily concludes that the one-loop mass shift of the re-scaled sG kink must be twice (in absolute value) the one-loop correction of the sG kink mass because the Hessian operator on the re-scaled sG kink is four times the Hessian operator on the sG kink. The estimation shown in the Table is in perfect concordance with this fact. Note, however, that the classical mass behaves in the opposite way: the classical mass of the re-scaled sG kink is half the classical mass of the proper sG kink.

The following entries in Table 1 come from other models. The unifying feature of all these other models is
the impossibility of applying the DHN formula because of the lack of information about the spectra of the Hessians. Therefore, all the one-loop corrections given in the Table above are computed by means of the asymptotic formula. The potential in the second row is the famous Razavy potential of molecular physics, see \cite{Razavy}. Since the potential of the physical pendulum gives birth to the (1+1)-dimensional sine-Gordon field theory, we use the Razavy potential of molecular physics to build another (1+1)-dimensional field theoretical model. The model in the third row is akin to the $\phi^6$ Lohe/Khare model \cite{Lohe}, \cite{Khare}. The difference is that in this case the minimum
at $\phi=0$ becomes a maximum and the vacuum orbit has only two points, see Table 1. In rows 4 and 5 we respectively show pure $\phi^8$ and $\phi^{16}$ models. Note that in these cases only implicit expressions of the kink solutions are available. Thus, there are no analytic expressions of the Hessians on these kinks available. The ability to compute the one-loop mass shift in this situation is a very strong aspect of the asymptotic approach: one is able to compute spectral information even if the analytical expression of the differential operator is not known !!. The required input is only the potential function $U(\phi)$. The model in the last row is very special. It has been derived from the deformation method, see \cite{BGLM}, and its moduli vacuum space has two points. There are thus six kink/antikink solutions
\[
{\rm cos}\left(\frac{{\rm arccos}({\rm tanh}x ) +(j-1)\pi}{3}\right) \quad , \quad j=1,2, \cdots , 6
\]
but only the one shown in Table 2 (and its antikink) interpolates between vacua belonging to the same point of the vacuum moduli space. In Table 2 we offer the pertinent data of all these models.

\begin{table}[ht]
\hspace{1cm}\begin{tabular}{|c|c|c|c|}
\hline
\rule[-0.5cm]{0.cm}{1.2cm} $U(\phi)$ & $\Phi_K(x)$ & Superpotential & $v^2$ \\ \hline
\rule[-0.5cm]{0.cm}{1.2cm} $1-\cos (2\phi)$ & $2 \arctan e^{2 x} $ & $\pm 2 \cos \phi$ & 4 \\ \hline
\rule[-0.5cm]{0.cm}{1.2cm} $\frac{1}{4}(\sinh^2\phi-1)^2$ & ${\rm arctanh} \frac{\tanh x}{\sqrt{2}}$ & $\frac{\pm 1}{2\sqrt{2}} [\frac{1}{2} \sinh 2\phi - 3\phi]$ & 4 \\ \hline
\rule[-0.5cm]{0.cm}{1.2cm} $\frac{1}{2}(\phi^2+1)(\phi^2-1)^2$ & $\frac{e^{2 \sqrt{2} x}-1}{\sqrt{1+6 e^{2 \sqrt{2} x}+e^{4 \sqrt{2} x}}}$  & $ \begin{array}{c} \frac{\pm 1}{8}[\phi\sqrt{1+\phi^2} (2\phi^2-3)-\\ -5 \,{\rm arcsinh}\, \phi] \end{array} $ & 8 \\ \hline
\rule[-0.5cm]{0.cm}{1.2cm} $\frac{1}{2}(\phi^4-1)^2$ & $-2\arctan \phi_K+\log\frac{1-\phi_K}{1+\phi_K}=4x$  & $\pm (\frac{1}{5} \phi^5-\phi)$ & 16 \\ \hline
$\frac{1}{2}(\phi^8-1)^2$ & $\begin{array}{c} 2\log \frac{1-\phi}{1+\phi} - \sqrt{2} \log \frac{1+\sqrt{2}\phi+\phi^2}{1-\sqrt{2}\phi+\phi^2}+ \\+2 \sqrt{2} \arctan \frac{1}{\phi^{2}}-4 \arctan \phi=16x \end{array}$   & $\pm (\frac{1}{9}\phi^9-\phi)$ & 64 \\ \hline \rule[-0.5cm]{0.cm}{1.2cm}
$\frac{8}{9}(\frac{1}{4}-\phi^2)^2(1-\phi^2)^2$ & ${\rm cos}\left(\frac{{\rm arccos}({\rm tanh}x ) +\pi}{3}\right)$   & $\pm (\frac{1}{3}\phi - \frac{5}{9} \phi^3+\frac{4}{15}\phi^5)$ & 4 \\ \hline
\end{tabular}
\caption{Kink solutions, superpotentials, and particle masses $v^2$ of the one-component scalar field theory models chosen above.}
\end{table}
We skip giving the specific parameters $m_d$ and $\gamma_d$ in each model, but its identification is not difficult in comparison with the physical parameters chosen in the Literature.

\subsection{One-loop kink mass corrections in two one-parametric families of models}

 We finally consider two families of real scalar field theory models with trigonometric and polynomial potentials depending on a real parameter $a$. The family of potentials in the first case is a family of double sine-Gordon potentials where the potential $U_{\rm sG}^{(2\pi)}$ is continuously deformed to the potential $U_{\rm sG}^{(\pi)}$. This example is a new stress test for our method because the quantum correction must interpolate between two known values. The second family of models is formed by a one-parametric family of generalized $\phi^6$ models with only two vacuum points that can be considered as a deformation of the $\phi^6$ Lohe/Khare model, where a third vacuum arises, see \cite{Lohe}-\cite{Khare}.

\subsubsection{One-loop kink mass shifts in a family of double sine-Gordon models}

The first family of models is characterized by the following family of potentials in (\ref{action}):
\[
U(\phi;a) = 1-(1-a)\cos \phi-a \cos (2\phi) \quad , \quad 0\leq a \leq 1 \, \, , \, \, a\in {\mathbb R} \quad .
\]
For fixed $a$, such that $0<a<1$, it is the celebrated double sine-Gordon model that has been studied
by many authors. We specifically mention Reference \cite{Mussardo} as the starting point for our calculations of one-loop kink mass shifts in the double sine-Gordon models; closely related models are considered in Reference \cite{Bazeia}.
Given $a$, $U(\phi;a)$ is a non-negative function that, when $a$ varies in the above range, interpolates between the sine-Gordon, $a=0$, and the re-scaled sine-Gordon, $a=1$, models:
\[
U(\phi;0)= 1-\cos \phi =U_{\rm sG}^{(2\pi)}(\phi) \hspace{1cm},\hspace{1cm} U(\phi;1)= 1-\cos 2\phi =U_{\rm sG}^{(\pi)}(\phi)
\]
For all the parameter values such that $0\leq a < 1$ the vacuum orbit is ${\cal M}=\{2\pi k\}$, $k\in \mathbb{Z}$. When $a$ becomes greater than $0$ but is strictly less than $1$, new relative minima (false vacua) emerge between each consecutive pair of vacua. At $a=1$ the new minima become absolute minima, the vacuum points of the re-scaled sine-Gordon model, see Figure 4 (left), such that the vacuum orbit becomes ${\cal M}=\{\pi k\}$, $k\in \mathbb{Z}$.
The kink solution of each member of the family with $a$ strictly less than $1$ is:
\begin{equation}
\phi_K(x;a)= -2 \arctan \frac{\sqrt{1+3a}}{\sqrt{1-a} \sinh (\sqrt{1+3a}x)} = \pi + 2 \arctan \frac{\sqrt{1-a} \sinh (\sqrt{1+3a}x)}{\sqrt{1+3a}}
\label{mixedsoliton}
\end{equation}
see again Figure 4 (middle left). The classical energy can be calculated from the superpotential
\[
W(\phi;a)=\frac{(a-1) \log \left(2 \sqrt{a} \cos \left(\frac{\phi}{2}\right)+\sqrt{2 a \cos (\phi)+a+1}\right)-2 \sqrt{a} \cos
   \left(\frac{\phi}{2}\right) \sqrt{2 a \cos (\phi)+a+1}}{\sqrt{a}}
 \]
 to find:
 \[
E(\phi_K(a))=\left|\frac{(a-1) \log \left(\frac{2 \sqrt{a}+\sqrt{3 a+1}}{\sqrt{3 a+1}-2 \sqrt{a}}\right)}{\sqrt{a}}-4 \sqrt{3 a+1}\right| \qquad .
\]
We can check that when $a=0$ the kink solution is $\phi_K(x;0)=4 \arctan e^x$: the kink of the sine-Gordon model. Its density energy is localized around one point. As the parameter $a$ approaches a value of $1$ the solution (\ref{mixedsoliton}) is formed by two separate kinks somehow related to the kink of the re-scaled sine-Gordon model. In this situation, the energy density shows two lumps of energy, see Figure 4 (middle right). The splitting of the two lumps is only appreciable near the value $a=1$. Note that for $a=0.9999$ the distance between the two lumps is less than eight spatial units, see Figure 4 (middle left). This is also manifest in the classical energy, see Figure 4 (right): if $a=0$ the classical energy is $8$ as it should be. For $a=1-\varepsilon$ we find almost $8$ again, meaning that we meet two re-scaled sG kinks well apart from each other. The consequence is that we must expect very close values to $a=1$ in order to observe the effect of the splitting in the kink mass quantum corrections.
\begin{figure}[h]
\centerline{\includegraphics[height=2.5cm]{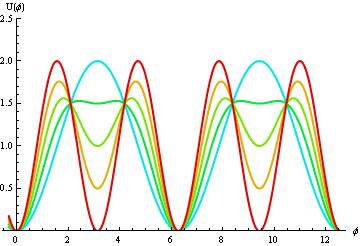}\hspace{0.3cm} \includegraphics[height=2.5cm]{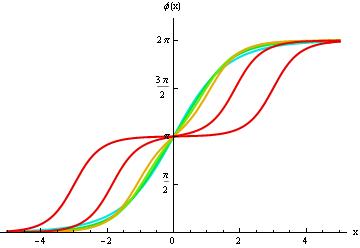}\hspace{0.3cm} \includegraphics[height=2.5cm]{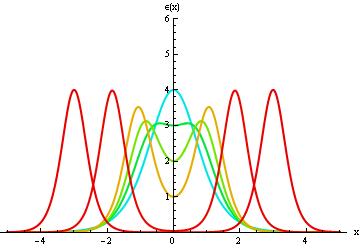} \hspace{0.3cm} \includegraphics[height=2.5cm]{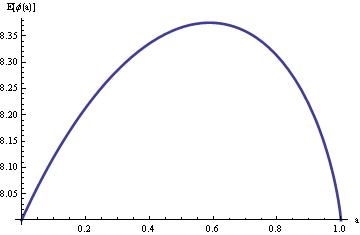}}
\caption{\small \textit{Graphical representation of the family of potentials for$a=0,0.25,0.5,0.75,1$ (left), kink solution profiles (middle left), kink energy densities (middle right) for $a=0,0.25,0.5,0.75,0.99,0.9999$ in several double sine-Gordon models. The kink classical energy as a function of $a$ (right).}}
\end{figure}
For completeness we give the particle masses and the quantum potential wells:
\[
v^2=1+3a \qquad , \qquad V(x)=\frac{4 (a-1) (3 a+1) \left((15 a+1) \cosh \left(2 \sqrt{3 a+1} x\right)-9 a+1\right)}{\left(-(a-1) \cosh \left(2 \sqrt{3a+1} x\right)+7 a+1\right)^2}
\]
which are not, in the new framework of this paper, necessary to compute the one-loop mass shifts.
We shall work the asymptotic approach to compute the one-loop quantum correction to the masses of the kinks  (\ref{mixedsoliton}) for several values of the parameter $a$. The Seeley coefficients are depicted for the values $a=0.1,0.2,0.3,\dots,0.9$ in Figure 5.

\begin{figure}[ht]
\hspace{6cm}\includegraphics[height=3.5cm]{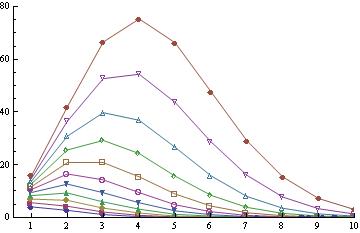}
\caption{\small Seeley coefficients in the family of double sine-Gordon models for $a=0.1,0.2,0.3,\dots,0.9$. The lines join the values of the coefficients associated with a specific value of $a$. The coefficients increase with increasing $a$.}
\end{figure}
The quantum corrections to the kink masses are given in Table 3, where the truncation order is also depicted. All these data are graphically represented in the attached Figure 5.
\begin{table}[ht]
\hspace{0.4cm}\begin{tabular}{|c|c|c|}
\hline
$a$ & $\Delta \tilde{E}/ \hbar m_d$ & $N$ \\ \hline
0.0 & -0.318321 & 20 \\ \hline
0.1 & -0.37569  & 7 \\ \hline
0.2 & -0.43009  & 7 \\ \hline
0.3 & -0.49231  & 7 \\ \hline
0.4 & -0.55860  & 9 \\ \hline
0.5 & -0.63262  & 10 \\ \hline
0.6 & -0.71320  & 10 \\ \hline
\end{tabular} \hspace{1cm}
\begin{tabular}{|c|c|c|}
\hline
$a$ & $\Delta \tilde{E}/ \hbar m_d$& $N$ \\ \hline
0.7     & -0.79229  & 10 \\ \hline
0.8     & -0.87735  & 10 \\ \hline
0.9     & -0.96929  & 10 \\ \hline
0.99    & -1.07836  & 12 \\ \hline
0.999   & -1.10794  & 15 \\ \hline
0.9999  & -1.11717  & 16 \\ \hline
0.99999 & -1.12197  & 17 \\ \hline
\end{tabular} \hspace{1cm}
\begin{tabular}{c}
\includegraphics[height=3.5cm]{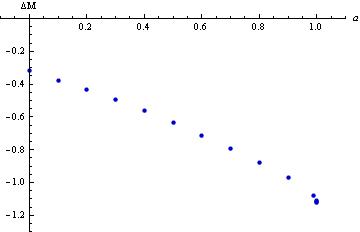}
\end{tabular}
\caption{\textit{The value of the quantum corrections to the kink masses in the family of double sine-Gordon models for several values of the parameter $a$ (left) and its graphical representation (right).}}
\end{table}
Bearing in mind that at the limit $a\rightarrow 1$ solution (\ref{mixedsoliton}) is formed by two solitons of the re-scaled sine-Gordon model, it is tempting to interpret the quantum correction $\Delta E$ at this limit as the double of the quantum correction of the soliton in the model mentioned. In particular, we recall that $\Delta E_{\rm sG}^{(\pi)} = -0.636894 \hbar m$, see the results in Table 1, and that the classical mass is twice the classical mass of the re-scaled sine-Gordon kink. We observe an approximated behaviour like this in Table 3, although the accumulation near the point $a=1$ does not allow us to reproduce the answer with complete precision.

\subsubsection{One-loop kink mass shifts in a family of generalized $\phi^6$ models}

The family of potentials in the action (\ref{action}) chosen now is
\[
U(\phi;a) = \frac{1}{2} (\phi^2 + a^2) (\phi^2-1)^2
\]
The vacuum orbit is ${\cal M}=\{-1,1\}$ and the kink solution that connects these vacuum points are
\begin{equation}
\phi_K(x;a)= \frac{a (-1+e^{2\sqrt{1+a^2}\,x})}{\sqrt{4e^{2\sqrt{1+a^2}\,x}+a^2 (1+e^{2\sqrt{1+a^2}\,x})^2}} \label{solutionphi6}
\end{equation}
From the family of superpotentials
\[
W(\phi;a)=\frac{1}{8} \left[a^2 \left(a^2+4\right) \log \left(2 \left(\sqrt{a^2+\phi^2}+\phi\right)\right)-\phi \sqrt{a^2+\phi^2} \left(a^2+2 \phi^2-4\right)\right]
\]
we read the classical energy of each member of the kink family:
\[
E(\phi_K(a))=-\frac{1}{4} \left(a^2-2\right) \sqrt{a^2+1}+\frac{1}{8}\left(a^2+4\right) a^2 \left(\log \frac{\sqrt{a^2+1}+1}{\sqrt{a^2+1}-1}\right) \quad .
\]
In Figure 6 some members of the family of potentials, some kink solutions, and their energy densities for several values of the parameter $a$ are depicted. A plot of the classical energy as a function of $a$ is also offered.
\begin{figure}[h]
\centerline{\includegraphics[height=2.5cm]{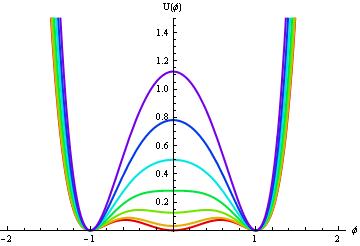}\hspace{0.3cm} \includegraphics[height=2.5cm]{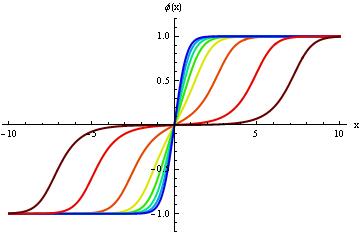} \hspace{0.3cm} \includegraphics[height=2.5cm]{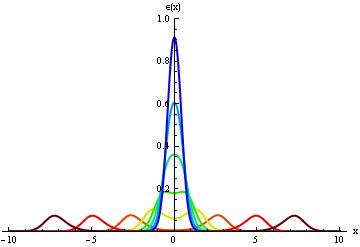}\hspace{0.3cm} \includegraphics[height=2.5cm]{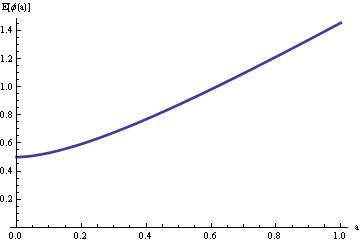}}
\caption{\small \textit{Graphical representation of potentials (left), kink solution profiles  (middle left), kink energy densities (middle right) for $a=0.001,0.01,0.25,0.5,0.75,1,1.25,1.5$ in some generalized $\phi^6$ models. The kink classical energy as a function of $a$ (right).}}
\end{figure}
The particle masses and the Schr$\ddot{\rm o}$dinger potential wells are in this case
\[
v^2=4(1+a^2) \qquad , \qquad V(x;a)=\frac{15 (4 a+1)^2}{\left(2 a \cosh \left(2 \sqrt{a^2+1} x\right)+2 a+1\right)^2}-\frac{6 \left(a^2+3\right) (4 a+1)}{2 a
   \cosh \left(2 \sqrt{a^2+1} x\right)+2 a+1} \quad .
\]
Now, the computational approach unveils the one-loop kink mass shifts shown in Table 4 for several values of the parameter $a$, with a truncation of $N=11$ terms. All these data are graphically represented in the attached Figure in Table 4.

\vspace{0.2cm}

\begin{table}[ht]
\hspace{2.5cm}\begin{tabular}{|c|c|}
\hline
$a$ & $\Delta \tilde{E}/ \hbar m_d$ \\ \hline
0.001 & -1.95304 \\ \hline
0.01 & -1.66578 \\ \hline
0.05 & -1.44663 \\ \hline
0.1 & -1.34908 \\ \hline
0.2 & -1.23924 \\ \hline
0.3 & -1.15669 \\ \hline
0.4 & -1.10130 \\ \hline
0.5 & -1.06892 \\ \hline
0.6 & -1.04909 \\ \hline
\end{tabular} \hspace{1cm}
\begin{tabular}{|c|c|}
\hline
$a$ & $\Delta \tilde{E}/ \hbar m_d$ \\ \hline
0.7 & -1.04611 \\ \hline
0.8 & -1.05411 \\ \hline
0.9 & -1.07127 \\ \hline
1.0 & -1.10077 \\ \hline
1.1 & -1.12710 \\ \hline
1.2 & -1.16278 \\ \hline
1.3 & -1.20309 \\ \hline
1.4 & -1.24630 \\ \hline
1.5 & -1.29266\\ \hline
\end{tabular} \hspace{1cm}
\begin{tabular}{c}
\includegraphics[height=4.8cm]{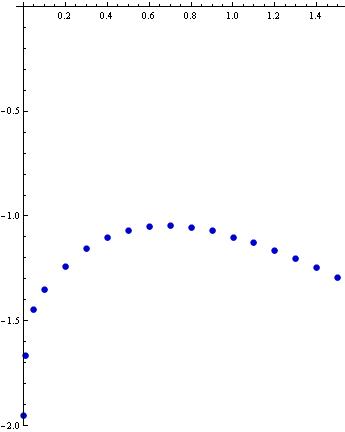}
\end{tabular}
\caption{\textit{Quantum corrections to the kink masses in some members of the generalized $\phi^6$ models for several values of the parameter $a$ (left) and its graphical representation (right).}}
\end{table}
We remark an interesting behavior. At the $a\rightarrow 0$ limit the potential reduces to $U(\phi,0)=\frac{1}{2} \phi^2 (\phi^2-1)^2$, i.e., the pure $\phi^6$ Lohe/Khare model; a new vacuum point $\phi=0$ arises, and the vacuum orbit becomes: ${\cal M}=\{-1,0,1\}$. The $\lim_{a\to 0}\phi_K(x;a)$ goes to a configuration where
the $\phi_K(a)$ kink solution splits into two kinks, one connecting the vacua $\phi=-1$ and $\phi=0$ and the other connecting $\phi=0$ and $\phi=1$, each one close to the kink of the Lohe/Khare model, see Figure 6. Note that as the parameter $a$ vanishes solutions (\ref{solutionphi6}) involve two separated lumps of energy. We observe in Table 4 that the quantum correction to these particular kinks tends to infinity. The reason for this stems from the fact that the vacua that are connected by this kink give rise to different meson masses: $\frac{\partial^2 U}{\partial \phi^2}[1]=4$ and $\frac{\partial^2 U}{\partial \phi^2}[0]=1$, see \cite{Alonso2002} and References therein to see a treatment of this problem.

\section{Summary and outlook}

In this paper and several earlier ones, \cite{Alonso2002}, \cite{Alonso2004}, \cite{Mateos2006}, \cite{Mateos2009},
we have elaborated a computational procedure to calculate one-loop kink mass shifts summarized in formula (\ref{numericalcorrection}) supplemented with a very sophisticated use of symbolic Mathematica software. The derivation of this formula is based on the high-temperature heat-kernel expansion plus the Mellin transform, thus ending in some series of asymptotic type. The old DHN approach to this problem led, after zero point renormalization by subtraction of the vacuum from the kink energy mode-by-mode, to formula (\ref{dhn}) which can be regularized either by a cutoff in the number of modes (\ref{dhncr}) or by means of the generalized zeta function procedure: (\ref{zolqc}). Except for the sine-Gordon and $\lambda\phi^4$ kinks these formulas are useless from a computational point of view because of the lack of information about the spectra of the differential operators
 governing the fluctuations around other interesting kinks. More precisely, what we have achieved in this work is
 some conceptual advance in the understanding of formula (\ref{numericalcorrection}) derived (and exploited) in the papers quoted above that hugely augmented the precision of the symbolic algorithm in our results on sine-Gordon and $\lambda\phi^4$ one-loop kink mass shifts. More importantly, the asymptotic approach has been applied to other systems supporting kinks considered in the Literature with a very high level of confidence and effectiveness. These results are even more remarkable because, to the best of our knowledge, there are no other ways of computing, even approximately, the one-loop kink mass shifts in these models. As a final remark on the advances reported here
we stress a very appealing property of our improved method: very few data are necessary for the computation of one-loop
kink mass shifts in this way. Only the potential and the vacua connected by the kink are needed. The number of terms kept
from the series is optimized in parallel to the choice of the integration range in the Mellin transform.

We plan to keep improving the asymptotic method by applying the procedure to more complex situations:

\begin{enumerate}

\item The first target would be the computation of one-loop kink mass shifts when the kinks interpolate among
vacua belonging to different points in the vacuum moduli space. In the $\phi^6$-model, this is always the case and the idea is to incorporate the improvements achieved here to the calculation of the $\phi^6$-kink mass shift performed in \cite{Alonso2002}. There are many more models carrying this type of kink susceptible to analysis within the framework of the improved asymptotic approach.

\item In kink-field theories with more than one scalar real field there are lots of very interesting kinks
that can have one- two- or $N$-components different from zero. The one-loop kink mass shifts of the celebrated
Montonen-Sarker-Trulinger-Bishop model \cite{Montonen}-\cite{STB} are computed in \cite{Alonso2002A}. Now we have the possibility of remarkably improving the precision of our results.

Another field theoretical system of this type is a two-component scalar field theory proposed by Bazeia et al. in \cite{BazeiaA}, who identified some isolated kinks. The solitary-wave solutions were unveiled by Shifman and Voloshin in \cite{Shifman}, whereas we related these kinks to critical trajectories in some integrable or quasi-integrable dynamical systems in \cite{Gonzalez}. The one-loop kink mass shifts were discussed in \cite{Alonso2004} and again we expect to get closer to the exact answer by applying the improved asymptotic approach.

\item BPS states in the $N=2$-supersymmetric $(1+1)$-dimensional Wess-Zumino model are kink solutions of first-order equations derived from holomorphic superpotentials. Derived in \cite{Fendley}, they are crucia elements in the classification of $N=2$-supersymmetric integrable systems. In \cite{Mateos2000} we discussed
    these special solitary waves from a real analytical point of view in order to compare them with the more traditional kinks arising in the models mentioned above. It seems natural to try the one-loop computation of mass shifts for these kinks restricting the theory to the bosonic sector: it is known that $N=2$ supersymmetry forces the
    quantum corrections to the BPS masses to be zero.

 \item Finally, we look forward, in a longer perspective, to improving the results on one-loop mass shifts of self-dual Nielsen-Olesen vortices achieved in \cite{Mateos2004} and \cite{Mateos2005} in the Abelian Higgs model. Simili modo, important improvements in the results on one-loop mass shifts of the planar self-dual topological solitons calculated in \cite{Mateos2008} in the semi-local Abelian Higgs model should be expected.

\end{enumerate}

\section*{ACKNOWLEDGEMENTS}

We warmly thank our collaborators in previous research on this topic W. Garcia Fuertes, M. Gonzalez Leon and M. de la Torre Mayado for illuminating conversations about different aspects of this subject.

We also gratefully acknowledge that this work has been partially financed
by the Spanish Ministerio de Educacion y Ciencia (DGICYT) under grant: FIS2009-10546.

\end{document}